\documentclass[lettersize,journal]{IEEEtran}
\usepackage{amsmath,amsfonts}
\usepackage{algorithmic}
\usepackage[ruled,linesnumbered]{algorithm2e}
\usepackage{array}
\usepackage{bm}
\usepackage[caption=false,font=normalsize,labelfont=sf,textfont=sf]{subfig}
\usepackage{textcomp}
\usepackage{stfloats}
\usepackage{url}
\usepackage{verbatim}
\usepackage{graphicx}
\usepackage{cite}
\usepackage{tabularx,booktabs}
\usepackage{multirow}
\usepackage{multicol}
\usepackage{colortbl} 
\usepackage{xcolor}
\usepackage{hyperref}
\usepackage{amssymb}
\usepackage{cite}
\usepackage{bbding}

\definecolor{lightblue}{RGB}{204,186,230}
\definecolor{citeblue}{RGB}{32,67,192}

\hypersetup{
	colorlinks=true,
	linkcolor=red,
	anchorcolor=blue,
	citecolor=green}

\newcommand{\bb}{\boldsymbol}
\newcommand{\tx}{\textcolor{black}}
\newcommand{\tc}{\textcolor{black}}

\newcommand{\e}{\emph{et al.}}

\newtheorem{theorem}{ \bf Theorem}
\newtheorem{lemma}{\bf Lemma}

\newtheorem{proposition}{\bf Proposition}
\newtheorem{definition}{\bf Definition}

\newtheorem{remark}{\bf Remark} 

\newcolumntype{L}[1]{>{\raggedright\arraybackslash}p{#1}}
\newcolumntype{C}[1]{>{\centering\arraybackslash}p{#1}}
\newcolumntype{R}[1]{>{\raggedleft\arraybackslash}p{#1}}

\makeatletter
\newcommand{\removelatexerror}{\let\@latex@error\@gobble}
\renewcommand{\maketag@@@}[1]{\hbox{\m@th\normalsize\normalfont#1}}%
\makeatother
\usepackage{makeidx}
\makeindex

\begin{document}
	\title{ A Weight-aware-based Multi-source Unsupervised Domain Adaptation Method for Human Motion Intention Recognition}
	\author{Xiao-Yin Liu,~\IEEEmembership{} Guotao Li*,~\IEEEmembership{} Xiao-Hu Zhou,~\IEEEmembership{} Xu Liang,~\IEEEmembership{} 
 Zeng-Guang Hou*,~\IEEEmembership{Fellow, IEEE}
		\thanks{This work is funded by the National Natural Science Foundation of China under (Grant 62473365, Grant U22A2056, and Grant 62373013), and the Beijing Natural Science Foundation under (Grant L222053, L232021, and L242101)(*Corresponding authors: Guotao Li and Zeng-Guang Hou).}
		\thanks{Xiao-Yin Liu, Guotao Li and Xiao-Hu Zhou are with the State Key Laboratory of Multimodal Artificial Intelligence Systems, Institute of Automation, Chinese Academy of Sciences, Beijing 100190, China, and also with the School of Artificial Intelligence, University of Chinese Academy of Sciences, Beijing 100049, China. (e-mail: liuxiaoyin2023@ia.ac.cn, guotao.li@ia.ac.cn, xiaohu.zhou@ia.ac.cn).}

  \thanks{Xu Liang is with the School of Automation and Intelligence, Beijing Jiaotong University, Beijing 100044, China (e-mail: liangxu2013@ia.ac.cn).}
  
		\thanks{Zeng-Guang Hou is with the State Key Laboratory of Multimodal Artificial Intelligence Systems, Institute of Automation, Chinese Academy of Sciences, Beijing 100190, China, also with the School of Artificial Intelligence, University of Chinese Academy of Sciences, Beijing 100049, China, and also with CASIA-MUST Joint Laboratory of Intelligence Science and Technology, Institute of Systems Engineering, Macau University of Science and Technology, Macao, China. (e-mail: zengguang.hou@ia.ac.cn).}
	}
	
	\markboth{}%
	{Shell \MakeLowercase{\textit{et al.}}: A Sample Article Using IEEEtran.cls for IEEE Journals}
	\maketitle
	
	\begin{abstract}
 Accurate recognition of human motion intention (HMI) is beneficial for exoskeleton robots to improve the wearing comfort level and achieve natural human-robot interaction. A classifier trained on labeled source subjects (domains) performs poorly on unlabeled target subject since the difference in individual motor characteristics. The unsupervised domain adaptation (UDA) method has become an effective way to this problem. However, the labeled data are collected from multiple source subjects that might be different not only from the target subject but also from each other. The current UDA methods for HMI recognition ignore the difference between each source subject, which reduces the classification accuracy. Therefore, this paper considers the differences between source subjects and develops a novel theory and algorithm for UDA to recognize HMI, where the margin disparity discrepancy (MDD) is extended to multi-source UDA theory and a novel weight-aware-based multi-source UDA algorithm (WMDD) is proposed. The source domain weight, which can be adjusted adaptively by the MDD between each source subject and target subject, is incorporated into UDA to measure the differences between source subjects. The developed multi-source UDA theory is theoretical and the generalization error on target subject is guaranteed. The theory can be transformed into an optimization problem for UDA, successfully bridging the gap between theory and algorithm. Moreover, a lightweight network is employed to guarantee the real-time of classification and the adversarial learning between feature generator and ensemble classifiers is utilized to further improve the generalization ability. The extensive experiments verify theoretical analysis and show that WMDD outperforms previous UDA methods on HMI recognition tasks. 
	\end{abstract}


	\section{Introduction} \label{sec:introduction}
\IEEEPARstart{S}{pinal} cord injury, stroke and neuromotor impairment have seriously reduced the quality of human life. Exoskeleton robots, including rehabilitation robots \cite{9968047,TRO,10401240} and power-assisted robots \cite{10075369,cybernetics_human_intention1}, have become one of the important ways to treat and recover these diseases. To improve the wearing comfort level and achieve natural human-robot interaction, human motion intention (HMI) has gained great concerns in exoskeleton robots and human-robot interaction \cite{molinaro2024estimating,10376386,cybernetics_human_intention2}. Accurate recognition of the HMI is beneficial for exoskeleton robots to improve the recovery effects. 

HMI refers to the fusion of various biological signals (such as electroencephalography, electromyogram, etc.) and non-biological signals (such as speed, torque, etc.) to identify movement patterns. Li \e \cite{10309911} provided a systematic review of the HMI recognition research for exoskeleton robots. The current methods for HMI can be divided into two types: model-based and model-free. The model-based methods involve the kinematics model \cite{8967017}, the dynamic model \cite{8974252}, and the musculoskeletal model \cite{7927483}. They establish the relationship between sensing signals and motion parameters to recognize HMI. The model-based methods are suitable for continuous HMI recognition. However, it consumes more time for the identification and calibration of parameters and is not robust when the task is complicated \cite{7927483}. 

The model-free methods map the sensing signals into target HMI directly without parameter identification \cite{9804816,yin2021sa,wang2021real}. The above model-free methods can achieve good performance on specific subject but fail to perform well on cross-subjects. Due to the difference in individual motor characteristics, such as kinematic properties, most current methods require labeling a large amount of data and training specific classifiers for each individual, which is burdensome \cite{10309911}. \tx{Unsupervised domain adaptation (UDA) is a common method to solve the above problem \cite{zhang2022gaussian,zhang2023ensemble,zhang2020unsupervised1,9976035}, where the source subjects (domains) data are labeled and the target subject (domain) data are unlabeled. Currently, the UDA methods have also been widely applied in many other fields, such as fault diagnosis \cite{10517281,10371381}
, physical information estimation \cite{10378867}
, classification tasks \cite{10529516,10174652}, and object identification \cite{10640069,10075484}. In these fields, the discrepancy between the source and target domains greatly reduces prediction performance. To alleviate the performance reduction caused by this discrepancy, single-source UDA (SUDA) utilizes the adversarial-based methods to align two domains \cite{ganin2016domain,long2017learning,saito2018maximum} or explores different metric learning schemes to minimize this divergence \cite{zhang2019bridging,zhang2020unsupervised,xiao2021dynamic}.}

In actual HMI recognition tasks, the labeled data are collected from multiple source subjects that might be different not only from the target subject but also from each other. However, the above UDA methods for HMI recognition don't consider the difference between the source subjects, which may result in a sub-optimal solution to the problem and hinder the improvement of classifier performance \cite{10029939}. \tx{Therefore, this paper considers the difference between the source subjects, and aims to propose a novel multi-source UDA (MUDA) method for HMI recognition so as to further improve the classification accuracy in the target subject.} In MUDA field, there are following two key challenges that need to be addressed.
\begin{enumerate}
    \item \textit{How to accurately and effectively measure the discrepancy between source domain (subject) and target domain (subject)}?
    \item Since the gap between the each source domain and target domain is different, \textit{how to integrate these differences to improve the classification accuracy and generalization ability}?
\end{enumerate}

\begin{figure*}
	\centering
	\includegraphics[width=1\textwidth]{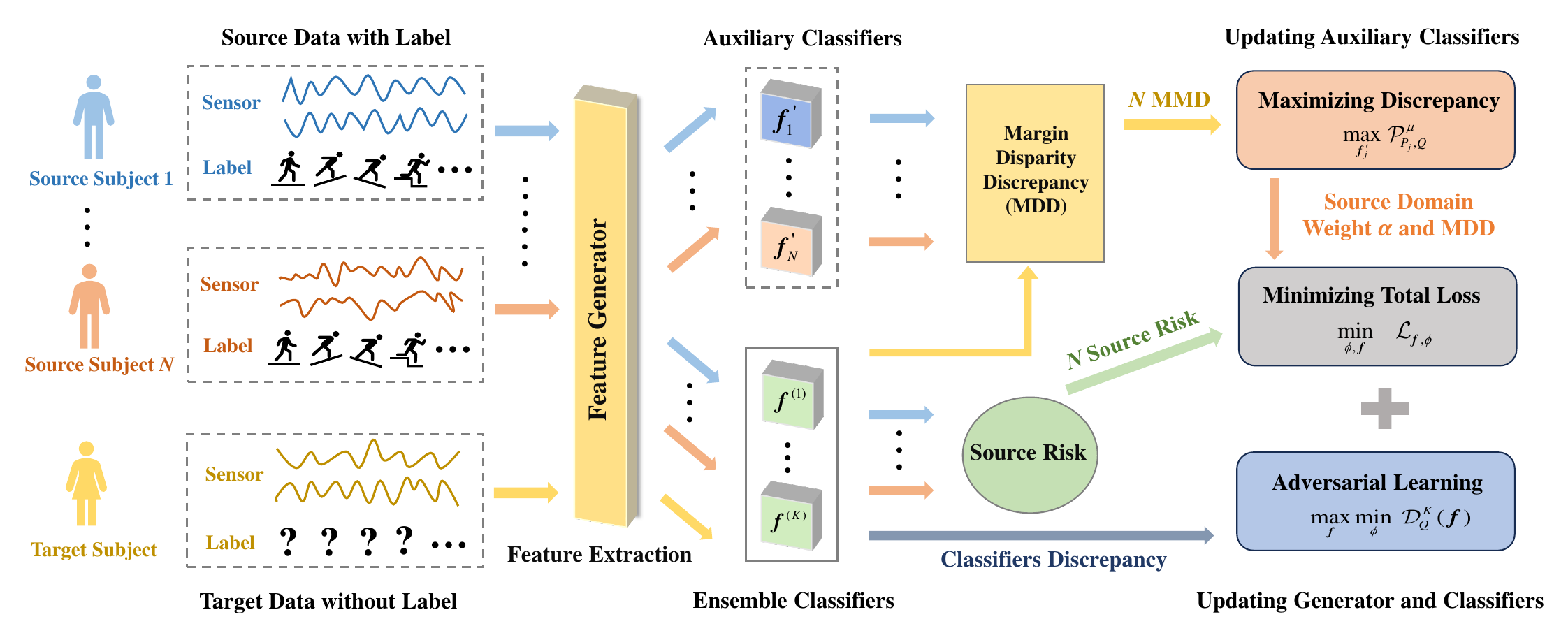}
	\caption{\tx{The overview of the weighted multi-source unsupervised domain adaptation (WMDD) method for human motion intention recognition. The training data are from $N$ source subjects with labels and one target subject without labels. The features of the sensor signal are extracted by feature generator. The final goal is to use feature generator and ensemble classifiers to classify the target subject intention accurately. Firstly, the auxiliary classifiers are optimized by maximizing the discrepancy, then the source domain weight can be achieved according to the estimated MDD. Secondly, the feature generator and classifiers are trained to minimize total loss. Finally, motivated by adversarial learning, the classifiers and generator are optimized by minimizing and maximizing classifiers discrepancy.}}. 
	\label{fig1}
\end{figure*}

For challenge 1,  Jensen Shannon Divergence \cite{ganin2015unsupervised}, Maximum Mean Discrepancy \cite{long2015learning} and Wasserstein Distance \cite{courty2017joint} are widely employed to measure the discrepancy between source and target domains in SUDA. Zhang \e \cite{zhang2019bridging} proposed Margin Disparity Discrepancy (MDD) method to measure this gap, which contains more generalization bound information. Zhang \e \cite{zhang2020unsupervised} aggregated the absolute margin violations in multi-class classification and proposed Multi-Class Scoring Disagreement Discrepancy based on MDD. However, the above discrepancy-based methods are based on the theory of SUDA. Further researches are required on the MUDA theory for related discrepancies.

\tc{For challenge 2, the current MUDA methods \cite{chen2024multi,wen2020domain,deng2023mixture,yan2017mind,yao2021multisource,wen2024maximum,9380556,hu2023swl}
introduced the source domain weight to integrate source domain differences and optimize the weighted sum loss of multiple sources to improve the accuracy of classification.} The source domain weight is mainly calculated through two types: solving optimization problem about domain weight \cite{chen2024multi,wen2020domain,deng2023mixture} and estimating domain weight based on the discrepancy or similarity between source and target domains \cite{yan2017mind,yao2021multisource,wen2024maximum,9380556,hu2023swl}. \tc{However, the first type may consume more computation cost when the parameter space of optimization problem is high. The second type may not better capture the discrepancy or similarity relationship between domains \cite{wang2023class} when source and target distributions are complex.} \tc{In addition, the theoretical foundations of weight-based MUDA methods are not complete, and there is still a gap between theory and algorithms.}

\tx{To solve the above problems, motivated by the margin disparity discrepancy \cite{zhang2019bridging} and distribution-weighted hypothesis \cite{blitzer2007learning}, we extend MDD in SUDA that contains more generalization information to MUDA field to accurately measure the discrepancy between source and target domains, and propose a novel weighted MUDA algorithm (WMDD), where the MDD is estimated by the auxiliary classifiers and source domain weight is adjusted by the estimated MDD. Moreover, the new generalization bound of MUDA is provided and proven, and the gap between theory and algorithm is analyzed in detail. In addition, the theoretical and empirical results validate that WMDD can learn diverse features and adapt well to unknown target subject in HMI recognition tasks.} Fig. \ref{fig1} shows the overview of the proposed WMDD method. The contributions of this paper are given below.
\begin{enumerate}
    \item \tc{A theory of MUDA is developed based on margin disparity discrepancy, and a novel weight-aware-based MUDA algorithm is proposed for HMI recognition.}
    \item The MDD between each source domain and target domain is estimated by auxiliary classifiers, which can adjust the source domain weight adaptively.
    \item \tc{A lightweight network is designed to guarantee the real-time of classification, and adversarial learning is utilized to further improve the generalization ability.}
\end{enumerate}

The framework of this paper is as follows: Section \ref{sec:Related work} introduces the related works about HMI recognition and MUDA. Section \ref{Theoretical Analysis} develops the new theory of MUDA, where the generalization bound for MUDA is proven. Section \ref{method} bridges the gap between MUDA theory and algorithm, and gives the detailed training steps of proposed method WMDD. Section \ref{sec:experiments} presents the performance of WMDD on HMI recognition tasks and verifies the advantages of the designed mechanism. Section \ref{Discussion} further discusses the multi-source UDA theory and algorithm. Section \ref{Conclusion} summarizes the entire work.

	\section{Related Work}\label{sec:Related work}
This section presents a brief overview of the literature in the area of human motion intention (HMI) recognition, the theory of domain adaptation, multi-source unsupervised domain adaptation (MUDA), and \tc{the differences between the proposed method of this paper and the previous related works.}

\subsection{Human Motion Intention Recognition}
\tc{Model-free methods based on deep networks, which can obtain higher-level features from sensor signals without domain-specific knowledge, have achieved great success in recent years, such as CNN \cite{9447716}, LSTM \cite{9174996}, CNN-BiLSTM \cite{sun2022continuous}. However, the model-free methods based on the above deep networks are difficult to perform well on cross-subjects problem due to individual differences.} Domain adaption is an effective method to cross-domain (subject) problems. Zhang \e \cite{zhang2020unsupervised1} incorporated an unsupervised cross-subject adaptation method to predict the HMI of the target subject without labels. \tx{Zhang \e \cite{zhang2022gaussian} proposed a novel non-adversarial cross-subject adaptation method (EDHKD) for HMI recognition. The ensemble diverse hypotheses method was designed to mitigate the cross-subject divergence \cite{zhang2023ensemble}. EDHKD is the state-of-the-art method for HMI using domain adaptation technology. It mitigates the cross-subject divergence by training feature generators to minimize the upper bound of the classification discrepancy, and maximizes the discrepancy among multiple feature generators to learn diverse and complete features.} However, the above methods ignore the difference between each source subject, essentially a single-source UDA method, which may cause a sub-optimal solution to the problem. Few multi-source UDA studies have focused on recognizing HMI. Therefore, this paper aims to propose a novel multi-source UDA method for HMI recognition. \textit{To the best of our knowledge, WMDD is the first work about multi-source UDA used for HMI recognition.}

\subsection{\tx{Unsupervised Domain Adaptation Theory}}
\tx{The study of domain adaptation theory is central of generalization error bound. They focus on providing the theoretical guarantee for prediction performance on target domain (subject).} Ben-David \e \cite{ben2006analysis,ben2010theory} conducted the pioneering theoretical works in domain adaptation field. \tx{They used the $\mathcal{H}\Delta\mathcal{H}$ (the symmetric difference hypothesis space for a hypothesis space $\mathcal{H}$) divergence to replace the traditional distribution discrepancies and overcame the difficulties in estimation from finite samples.} Mansour \e \cite{mansour2009domain} extended the zero-one loss of \cite{ben2006analysis} to the general loss function of binary classification and developed a generalization theory. Kuroki \e \cite{kuroki2019unsupervised} proposed a more tractable source-guided discrepancy by fixing the hypothesis of \cite{mansour2009domain} to the ideal source minimizer. Zhang \e \cite{zhang2019bridging} extended the theory of \cite{ben2010theory} to multiple classes by introducing margin disparity discrepancy (MDD), which can characterize the difference of the multi-class scoring hypothesis. Zhang \e \cite{zhang2020unsupervised} proposed multi-class scoring disagreement divergence based on MDD \cite{zhang2019bridging}, that can characterize element-wise disagreements of multi-class scoring hypotheses by aggregating violations of absolute margin.

\subsection{\tx{Multi-source Unsupervised Domain Adaptation Algorithm} }
\tx{The current MUDA methods focused on aligning the distribution of each pair of source and target to reduce their domain shift by minimizing the combined discrepancy between source and target domains. The source domain weight was used to integrate source domain differences, and mainly calculated through solving related optimization problem \cite{chen2024multi,wen2020domain} or calculating the discrepancy or similarity between source and target domains \cite{yan2017mind,yao2021multisource,wen2024maximum,9380556,hu2023swl}. \tc{For the first-type methods, Chen \e \cite{chen2024multi} used Limited memory Broyden-Fletcher-Goldfarb-Shanno algorithm to solve domain weight,} and optimized networks by minimizing the source mixture loss and the Pearson divergence. Wen \e \cite{wen2020domain} solved domain weight through reformulating complex optimization problem, and optimized networks by minimizing source loss and discrepancy.} 

\tc{For the second-type methods, Yan \e \cite{yan2017mind} calculated domain weight based on class prior distribution, and proposed weighted domain adaption network to alleviate the effect of class weight bias.}
Yao \e \cite{yao2021multisource} quantified the importance of different source domains and aligned the source and target distributions by minimizing maximum mean discrepancy. Wen \e \cite{wen2024maximum} proposed a maximum likelihood weight estimation approach to estimate source weight function, which matches the source of relevant part to target domain. Zuo \e \cite{9380556} solved domain weight through calculating the frequencies of target domain belonging to each source domain, and  paid more attention on the source domains with higher similarities. Hu \e \cite{hu2023swl} introduced weight allocator to map the classification loss and domain discrimination loss of each sample to its weight in weighted domain alignment loss. 

\tc{However, the first-type methods would consume high computation cost when the parameter space is large since the weights are constantly optimized in training process. The above second-type methods may fail to better capture the relationship between different domains when the domain distributions are complex. The proposed method of this paper belongs to the second-type method. 
We determine the domain weight through the estimated MDD with less computation cost, which can be adaptively adjusted to better describe the correlation between each source domain and target domain.}

\subsection{\tc{The Proposed Method} }
\tc{The proposed method of this paper is different from the above multi-source domain adaptation methods that utilize source domain weight in the below three aspects:}

\textbf{1) Weight determination:} The source domain weight is determined based on the estimated margin disparity discrepancy, instead of class prior distribution \cite{yan2017mind}, maximum likelihood estimation \cite{wen2024maximum} or sample correction \cite{9380556,hu2023swl}. 
Compared with the domain weight optimized constantly by solving unconstrained problem \cite{chen2024multi}, \tc{the proposed method can reduce the computation cost and enhance algorithm stability.}

\textbf{2) Theoretical guarantee:} There are fewer complete theoretical guarantees and analysis in the above MUDA methods \cite{chen2024multi,wen2020domain,deng2023mixture,yan2017mind,yao2021multisource,wen2024maximum,9380556,hu2023swl}. We provide the complete theoretical guarantee for MUDA based on MDD theory. The theory of WMDD is based on each single source distribution instead of the whole multi-source class-conditionals \cite{yao2021multisource}. \tc{The theory of the proposed method of this paper fully considers the differences between each source domain, which is beneficial for improving classification accuracy.}

\tx{\textbf{3) Discrepancy measurement and real-time:} The distribution disparity metric is margin disparity discrepancy that contains more generalization bound information, rather than maximum mean discrepancy \cite{yao2021multisource} or Pearson divergence \cite{chen2024multi}. A lightweight network is employed to guarantee the real-time of classification, instead of using multiple complex networks \cite{wen2020domain} that is difficult to guarantee the real-time.}

	\section{Theoretical Analysis of Multi-source Unsupervised Domain Adaptation}\label{Theoretical Analysis}
In this section, we give the basic notations and generalization bound for multi-source UDA based on the theory of single-source UDA. The theoretical analysis follows the previous work \cite{zhang2019bridging}, where margin disparity discrepancy was used to measure the generalization bound.

\begin{figure}
	\centering
	\includegraphics[width=0.5\textwidth]{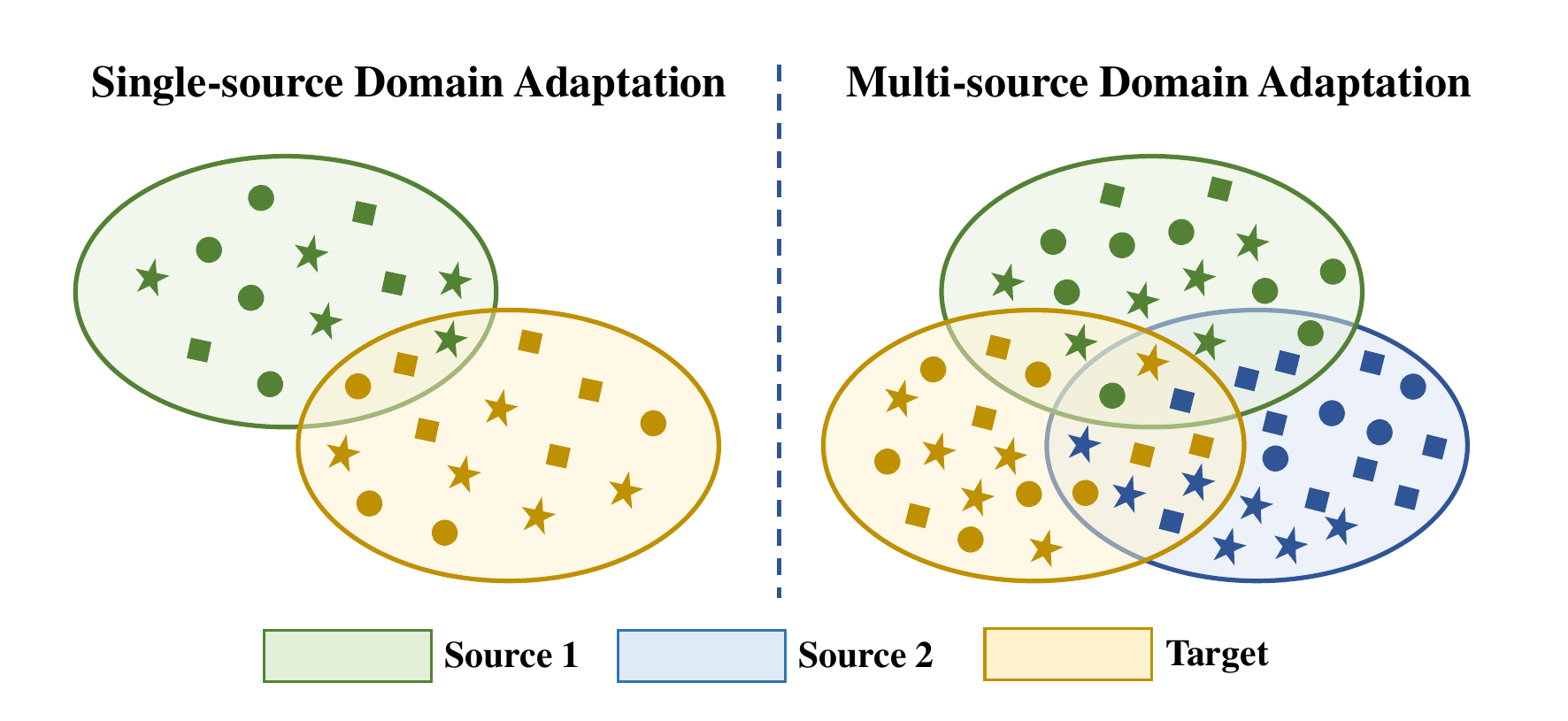}
	\caption{\tx{The comparison between the single-source unsupervised domain adaptation (SUDA) and multi-source unsupervised domain adaptation (MUDA), where different shapes represent different data categories and different colors denotes different domains. The source and target distributions of SUDA are not matched well. For MUDA, the target distribution hardly matches all subject distributions, and the discrepancy between the target distribution and each subject distribution might be different.}}. 
	\label{fig2}
\end{figure}

\subsection{Problem Formulation}
Fig. \ref{fig2} compares single-source unsupervised domain adaptation (SUDA) and multi-source unsupervised domain adaptation (MUDA). In MUDA, we consider $\mathcal{C}$ class, $N$ source domains $\{\mathcal{D}_{sj}\}_{j=1}^N$ and one target domain $\mathcal{D}_{t}$ problem. \tc{There are two different but related distributions over $\mathcal{X} \times \mathcal{Y}$, namely source distribution $P$ and target distribution $Q$. The term $\mathcal{X}$ denotes input space and $\mathcal{Y}$ denotes output space, where $\mathcal{Y}$ is $\{0, 1\}$ in binary classification.}

\tc{The learner of UDA is trained on the labeled data $\mathcal{D}_{sj}=\{(\bb{x}_i^{sj},y_i^{sj})\}_{i=1}^{|\mathcal{D}_{sj}|}$ and the unlabeled data $\mathcal{D}_{t}=\{\bb{x}_i^{t}\}_{i=1}^{|\mathcal{D}_{t}|}$, where the labeled data are sampled from source distribution $P_j$ and the unlabeled data are sampled from target distribution $Q$.}
The goal of MUDA is to learn the hypothesis space $\mathcal{H}$ of labeling function $h: \mathcal{X} \rightarrow  \mathcal{Y}$ to minimize expected target error
\begin{equation}
	\label{1}
	\mathcal{L}_Q(h)=\mathbb{E}_{(\bb{x}, y) \sim Q} L\left[h(\bb{x}), y\right],
\end{equation}
where $L$ is the loss function. Ben-David \e \cite{ben2010theory} used zero one loss of the form $\mathbb{I}[h(\bb{x})\ne y]$ to represent $L$, where $\mathbb{I}$ is the indicator function. \tx{Following \cite{zhang2020unsupervised}, we consider the hypothesis space $\mathcal{F}$ that contains scoring function $\bb{f}: \mathcal{X} \rightarrow \mathbb{R}^{\mathcal{C}}$ which outputs the prediction confidence.} Then the labeling function can be induced
\begin{equation}
	\label{2}
	h_{\bb{f}}(\bb{x})=\arg \max_{y\in \mathcal{Y}} f_{y}(\bb{x}),
\end{equation}
where $f_y$ denotes the $y$-th component of vector function $\bb{f}$. \tx{Zhang \e \cite{zhang2019bridging} defined the \emph{margin} of $\bb{f}$ for labeled sample $(x, y)$ as $\omega_{\bb{f}}(\bb{x},y)=[f_y(\bb{x})-\max_{y'\neq y}f_{y'}(\bb{x})]/2$.} Then, the margin loss of $\bb{f}$ can be denoted as
\begin{equation}
	\label{3}
	\mathcal{L}_{P}^{\mu}(\bb{f})=\mathbb{E}_{(\bb{x},y)\sim P}\Big\{ \mathbb{W}_\mu\big[\omega_{\boldsymbol{f}}(\bb{x},y)\big]\Big\},
\end{equation}
where $\mathbb{W}_\mu$ is ramp loss defined as
\begin{equation}
	\label{4}
	\mathbb{W}_\mu(v):= \begin{cases}0, & \mu \leq v \\ 1-v / \mu, & 0<v<\mu. \\ 1, & v \leq 0\end{cases}
\end{equation}

\tc{Margin disparity discrepancy (MDD) \cite{zhang2019bridging}, containing more generalization information, is used to measure the distribution divergence between $P$ and $Q$, which is defined as}
\begin{equation}
	\label{5}
	d^{\mu}_{\bb{f}}\left(P, Q\right)=\sup _{\boldsymbol{f}^{\prime} \in \mathcal{F}}\Big\{\mathbb{E}_{Q}\big[ \mathbb{W}_\mu\left(\omega_{\boldsymbol{f}^{\prime}}\right)\big]-\mathbb{E}_{P} \big[\mathbb{W}_\mu\left(\omega_{\boldsymbol{f}^{\prime}}\right)\big]\Big\},
\end{equation}
where $\mathbb{E}_{Q}[ \mathbb{W}_\mu\left(\omega_{\boldsymbol{f}^{\prime}}\right)] = \mathbb{E}_{\bb{x}\sim Q}\{ \mathbb{W}_\mu\left[\omega_{\boldsymbol{f}^{\prime}}(\bb{x},h_{\bb{f}}(\bb{x}))\right]\}$. Thus the gap between the $j$-th source distribution $P_j$ and target distribution $Q$ for MUDA can be defined as $d^{\mu}_{\bb{f}}(P_j,Q)$. For single-source UDA, we have the following cross-domain generalization bound between source distribution $P$ and target distribution $Q$:
\begin{theorem}
	\label{the_3}
	Fix $\mu>0$. For any scoring function $\bb{f} \in \mathcal{F}$,
	\begin{equation}
		\label{6}
		\mathcal{L}_Q\left(h_{\bb{f}}\right) \leq \mathcal{L}_{{P}}^{\mu}\left(\bb{f}\right)+d^\mu_{\bb{f}}\left(P, Q\right)+\lambda,
	\end{equation}
\end{theorem}
where the constant $\lambda$ is the ideal combined margin loss, defined as $\lambda=\min_{\bb{f}\in\mathcal{F}}\{\mathcal{L}_{{P}}^{\mu}(\bb{f})+\mathcal{L}_{Q}^{\mu}(\bb{f})\}$. 

The proof for Theorem \ref{the_3} can be found in Appendix \ref{proof_3}. In the following section, we give the cross-domain generalization bound for MUDA based on Theorem \ref{the_3}.
	\subsection{Generalization Bound for MUDA}
In MUDA, some source domains, which are more relevant to the target domain than others, might be more important to classify. Therefore, we use domain weight $\alpha_j\geq0$ to describe this and give the following theorem.

\begin{theorem}
	\label{the_1}
	Fix $\mu>0$ and the $N$ source domain datasets $\{\bb{X}_j,\bb{Y}_j\}_{j=1}^N$. For any scoring function $\bb{f} \in \mathcal{F}$ and $\bb{\alpha} \in \Delta=\{\boldsymbol{\alpha}: \alpha_j \geq 0, \sum_j \alpha_j=1\}$, the following holds:
	\begin{equation}
		\label{7}
			\mathcal{L}_Q\left(h_{\bb{f}}\right) \leq \sum_{j=1}^N \alpha_j\left(\mathcal{L}_{{P}_j}^{\mu}\left(\bb{f}\right)+d^\mu_{\bb{f}}\left(P_j, Q\right)\right)+\beta,
	\end{equation}
\end{theorem}
where $\bb{\alpha}$ is the weight of $N$ source domain, $\beta$ is the weighted margin loss combination, $\beta=\min_{\bb{f}\in\mathcal{F}}\{\sum_{j=1}^N \alpha_j\mathcal{L}_{{P}_j}^{\mu}(\bb{f})+\mathcal{L}_{Q}^{\mu}(\bb{f})\}$, which can be reduced to a rather small value if the hypothesis space is enough.

\begin{remark}
    The proof of Theorem \ref{the_1} can be found in Appendix \ref{proof_1}. \tx{Given the fixed $\beta$, the target error (generalization error) $\mathcal{L}_Q\left(h_{\bb{f}}\right)$ is determined by the distribution discrepancy $d^\mu_{\bb{f}}(P_j, Q)$ and the expected loss $\mathcal{L}_{{P}_j}^{\mu}(\bb{f})$. In Eq. \eqref{7}, the discrepancy $d^\mu_{\bb{f}}(P_j, Q)$ bounds the performance gap caused by domain distribution shift}. The smaller $\mathcal{L}_{{P}_j}^{\mu}(\bb{f})$ indicates better performance of scoring function $\bb{f}$ on the $j$-th source domain. 
\end{remark}

\tc{Note that, the above theorem only considers the ideal and complete distribution. However, the collected data fails to cover the entire distribution space in actual learning. Therefore, we further consider the sampling error and give the empirical form of Theorem \ref{the_1}.} First, Rademacher complexity that is widely used in the generalization theory is introduced to measure the richness of a certain hypothesis space \cite{mohri2012new,zhang2019bridging}. The definition is given below:

\begin{definition}[Empirical Rademacher complexity]\label{def_1}
	Let $\mathcal{G}$ be a family of functions mapping from $\mathcal{Z}$ to $\mathbb{R}$ and $\widehat{\mathcal{S}}=\left\{z_1, \ldots, z_{T}\right\}$ be a fixed sample of size $T$ drawn from the distribution $\mathcal{S}$ over $\mathcal{Z}$. The empirical Rademacher complexity of $\mathcal{G}$ for sample $\widehat{\mathcal{S}}$ is defined as
	\begin{equation}
		\label{8}
		\widehat{\Re}_{\widehat{\mathcal{S}}}(\mathcal{G})= \mathbb{E}_{\bb{\sigma}} \bigg[\sup _{g \in \mathcal{G}} \frac{1}{T}\sum_{t=1}^T \sigma_t g\left(z_t\right)\bigg],
	\end{equation}
\end{definition}
where $\bb{\sigma}=(\sigma_1,...,\sigma_{T})$ are independent uniform random variables taking values in $\{-1,+1\}$.

\begin{definition}[Induced Scoring Function Families]\label{def_2}
Given for a space $\mathcal{F}$ of scoring function $\bb{f}$, two induced scoring function families $\Omega_1(\mathcal{F})$ and $\Omega_2(\mathcal{F})$ are defined as
\begin{equation}\label{9}
    \begin{aligned}
	&\Omega_1(\mathcal{F}) = \Big\{(\bb{x},y)\rightarrow \bb{f}_y(\bb{x})~\big|~\bb{f}\in\mathcal{F}\Big\}, \\
        &\Omega_2(\mathcal{F}) = \Big\{\bb{x}\rightarrow \bb{f'}_{h_{\bb{f}}(\bb{x})}(\bb{x})~\big|~\bb{f}\in\mathcal{F},\bb{f'}\in\mathcal{F}\Big\}, 
  \end{aligned}
\end{equation}
\tc{where $\Omega_1(\mathcal{F})$ and $\Omega_2(\mathcal{F})$ can be regarded as the generated space based on the space $\mathcal{F}$.} Then, according to the above two definitions, the generalization bound of multi-source UDA can be given in the below theorem.
\end{definition}

\begin{theorem}
	\label{the_2}
	\tx{Let $\widehat{P}_j$ and $\widehat{Q}$ be the corresponding empirical distributions of ${P}_j$ and ${Q}$. $\widehat{\mathcal{D}}_{sj}$ and $\widehat{\mathcal{D}}_t$ are the empirical datasets sampled from samples $\mathcal{D}_{sj}$ and $\mathcal{D}_t$ with the size $m$, respectively.} \tc{Then, for any $\delta>0$, any $\bb{\alpha} \in \Delta=\{\boldsymbol{\alpha}: \alpha_j \geq 0, \sum_j \alpha_j=1\}$ and any scoring function $\bb{f} \in \mathcal{F}$, with probability at least $1-3\delta$, the following holds,}
 \begin{equation*}
		\begin{aligned}
			\mathcal{L}_Q\left(h_{\bb{f}}\right) \leq &\sum_{j=1}^N \alpha_j\Big\{\mathcal{L}_{{\widehat{P}}_j}^{\mu}\left(\bb{f}\right)+d^\mu_{\bb{f}}\left(\widehat{P}_j, \widehat{Q}\right)\Big\}+9\sqrt{\frac{\log (2 / \delta)}{2m}}\\
			&+\frac{\mathcal{C}}{\mu}\Big\{{\widehat{\mathfrak{R}}}_{\widehat{\mathcal{D}}_s}(\Omega_1(\mathcal{F}))+{\widehat{\mathfrak{R}}}_{\widehat{\mathcal{D}}}(\Omega_2(\mathcal{F}))\Big\}+\beta,
		\end{aligned}
\end{equation*}
where $\mathcal{C}$ denotes the number of classes, $\beta$ is constant independent of $\bb{f}$, $\widehat{\mathcal{D}}_s$ and $\widehat{\mathcal{D}}$ are datasets sampled from the mixture distribution $\sum_{j=1}^{N}\alpha_jP_j$ and $Q+\sum_{j=1}^{N}\alpha_jP_j$, respectively.
\end{theorem}

\begin{remark}
     The proof of Theorem \ref{the_2} can be found in Appendix \ref{proof_2}. \tc{Theorem \ref{the_2} indicates that the expected error on target domain $\mathcal{L}_Q\left(h_{\bb{f}}\right)$ is bounded by four parts, including empirical margin error on source domain $\mathcal{L}_{{\widehat{P}}_j}^{\mu}$, empirical distribution gap $d^\mu_{\bb{f}}(\widehat{P}_j, \widehat{S})$, the ideal error $\beta$ and rademacher complexity terms.} \tx{The better generalization ability can be achieved by selecting an appropriate margin $\mu$ for margin loss. Since the term $\beta$ is to take the minimum value in space $\mathcal{F}$, the upper bound of target error becomes tighter for relatively larger $\mu$ and rich hypothesis space $\mathcal{F}$. However, this bound cannot reach an acceptable small value for too large $\mu$.}  
\end{remark}
	\section{Connecting Theory and Algorithm} \label{method}
\tc{Based on the upper bound of the expected error given in Theorem \ref{the_2}, a novel theoretical-ground multi-source UDA algorithm is proposed in this section. We bridge theory and algorithm for MUDA, and elaborate on how to use lightweight networks to approximate the various terms of the above theory. Then, the detailed training steps of WMDD and the algorithm description are provided in this part.} 

\subsection{Theoretical Optimization Goal}
\tx{The goal of optimization is to minimize the target error $\mathcal{L}_Q(h_{\bb{f}})$ in Theorem \ref{the_2}, that is minimizing the upper bound of generalization error. \tc{Given hypothesis space $\mathcal{F}$ and samples $\mathcal{D}_{sj}$ and $\mathcal{D}_t$ and margin $\mu$, the ideal error $\beta$ and complexity terms are assumed to be fixed.} Therefore, the optimization goal can be written as:}
\begin{equation}
	\label{12}
	\min_{\bb{f}}\quad \sum_{j=1}^N \alpha_j\Big\{\mathcal{L}_{{\widehat{P}}_j}^{\mu}\left(\bb{f}\right)+d^\mu_{\bb{f}}\left(\widehat{P}_j, \widehat{Q}\right)\Big\},
\end{equation}
\tx{where $\mathcal{L}_{{\widehat{P}}_j}^{\mu}(\bb{f})$ denotes the expected error over the $j$-th source domains, and $d^\mu_{\bb{f}}(\widehat{P}_j, \widehat{Q})$ is the margin disparity discrepancy between the $j$-th source and target distributions. Then by substituting Eqs. \eqref{3} and \eqref{5} into Eq. \eqref{12}, we have}
\begin{equation}
	\label{13}
	\begin{aligned}
		&\min_{\bb{f}}\quad \sum_{j=1}^N \alpha_j\Big\{\underbrace{\mathbb{E}_{(\bb{x},y)\sim \widehat{P}_j}\big( \mathbb{W}_\mu\big[\omega_{\boldsymbol{f}}(\bb{x},y)\big]\big)}_{\tx{\text{Expected loss in $j$-th source: }\mathcal{L}_{{\widehat{P}}_j}^{\mu}\left(\bb{f}\right)}}\\
		+&\underbrace{\max _{\boldsymbol{f'}_j \in \mathcal{F}}\big(\mathbb{E}_{\widehat{Q}}\big[ \mathbb{W}_\mu\big(\omega_{\boldsymbol{f'}_j}\big)\big]-\mathbb{E}_{\widehat{P}_j} \big[\mathbb{W}_\mu\big(\omega_{\boldsymbol{f'}_j}\big)\big]\big)}_{\tx{\text{Margin disparity discrepancy: }d^\mu_{\bb{f}}\left(\widehat{P}_j, \widehat{Q}\right)}}\Big\}.
	\end{aligned}
\end{equation}

\tx{The above optimization problem is a minmax game problem. The form of optimization can be reformed as:}
\begin{equation}
	\label{14}
	\begin{aligned}
		&\min_{\bb{f}\in\mathcal{F}} \quad \sum_{j=1}^N \alpha_j\Big\{\mathcal{L}_{{\widehat{P}}_j}^{\mu}\left(\bb{f}\right)+\mathcal{P}_{\widehat{P}_j, \widehat{Q}}^\mu\left({\bb{f}},\bb{f'}_j\right)\Big\},\\
		&\text{ s.t.}\quad\text{  } \bb{f'}_j\in \mathop{\arg\max}\limits_{\bb{f'}\in\mathcal{F}}\quad \mathcal{P}_{\widehat{P}_j, \widehat{Q}}^\mu\left({\bb{f}},\bb{f'}\right),
	\end{aligned}
\end{equation}
where the maximization of $\mathcal{P}_{\widehat{P}_j, \widehat{Q}}^\mu=\mathbb{E}_{\widehat{Q}}\big(\mathbb{W}_\mu\big)-\mathbb{E}_{\widehat{P}_j} \big(\mathbb{W}_\mu\big)$ is the estimated MDD. \tx{In this optimization problem, the auxiliary function $\bb{f'}_j$ is first updated, and then $\bb{f}$ is updated based on the result of $\bb{f'}_j$.} The key of the next step is to determine the value of $\bb{\alpha}$. Intuitively, the source domain with a small gap to the target domain should have a large weight. Therefore, $\bb{\alpha}$ can be defined as
\begin{equation}
	\label{15}
	\alpha_j=\frac{\exp{\big\{-\big|\mathcal{P}_{\widehat{P}_j, \widehat{Q}}^\mu\left({\bb{f}},\bb{f'}_j\right)\big|\big\}}}{\sum_{j=1}^{N}\exp{\big\{-\big|\mathcal{P}_{\widehat{P}_j, \widehat{Q}}^\mu\left({\bb{f}},\bb{f'}_j\right)\big|\big\}}}.
\end{equation}

The source domain weight $\bb{\alpha}$ can be adjusted dynamically by the estimated MDD, that is weight-aware-based. Then combining Eqs. \eqref{14} and \eqref{15}, we can achieve final theoretical optimization goal of MUDA.

\subsection{Bridging Theory and Algorithm}
There is a gap between theory and algorithm since it is impractical to solve the above optimization problem directly. \tx{Firstly, due to the complexity of input space $\mathcal{X}$, we utilize the feature extractor $\phi(\cdot)$ to map the high input space $\mathcal{X}$ into low feature space $\Omega = \{\phi(\bb{x})~|~\bb{x}\in \mathcal{X}\}$.} Then label function can be again defined as $h_{\bb{f}}: \phi(\bb{x})\rightarrow \arg\max_{y\in\mathcal{Y}}f_{y}(\phi(\bb{x}))$.

Secondly, \tc{since the ramp loss used in the margin loss is prone to causing gradient vanishing, it is challenging to optimize Eq. \eqref{14} using stochastic gradient descent (SGD) \cite{zhang2020unsupervised}.} Therefore, \tc{we aim to find a surrogate function, denoted as $\mathcal{T}(\bb{f}(\phi(\bb{x})),y)$, for the margin loss. This surrogate function should be easily trainable using SGD while retaining the key properties of the margin loss.} Then, the two terms in Eq. \eqref{14} can be written as
\begin{equation}
	\label{16}
	\begin{aligned}
		\mathcal{L}_{{\widehat{P}}_j}^{\mu}\left(\bb{f}\right)=& \mathbb{E}_{(\bb{x},y)\sim \widehat{P}_j} \Big[\mathcal{T}\big(\bb{f}(\phi(\bb{x})),y\big)\Big]\\
		\mathcal{P}_{\widehat{P}_j, \widehat{Q}}^\mu\left({\bb{f}},\bb{f'}_j\right)=&\mathbb{E}_{\bb{x}\sim\widehat{Q}}\left[\mathcal{T}\big(\bb{f}(\phi(\bb{x})), h_{\bb{f'}_j}(\phi(\bb{x}))\big)\right]-\\
		&\gamma\mathbb{E}_{\bb{x}\sim\widehat{P}_j} \left[\mathcal{T}\big(\bb{f}(\phi(\bb{x})), h_{\bb{f'}_j}(\phi(\bb{x}))\big)\right],
	\end{aligned}
\end{equation}
where $\gamma$ is used to attain margin $\mu$ of margin loss similar to \cite{zhang2019bridging}. The cross-entropy loss is used to represent surrogate function $\mathcal{T}(\cdot,\cdot)$ following \cite{zhang2019bridging,zhang2020unsupervised}. On source subjects, the standard cross-entropy loss is employed, that is
\begin{equation}
	\label{17}
	\begin{aligned}
		\mathcal{T}\big(\bb{f}(\phi(\bb{x})),y\big) &= -\log \Big[ \Theta_y\big(\bb{f}(\phi(\bb{x}))\big)\Big]\\
		\mathcal{T}\big(\bb{f}(\phi(\bb{x})), h_{\bb{f'}_j}(\phi(\bb{x}))\big)&=-\log \Big[ \Theta_{h_{\bb{f'}_j}}\big(\bb{f}(\phi(\bb{x}))\big)\Big],
	\end{aligned}
\end{equation}
where $\Theta_y(\bb{f})$ is the softmax function, denoted as $\Theta_y(\bb{f}) = \text{exp}({\bb{f}_y})/\sum_{j=1}^{N}{\text{exp}({\bb{f}_j})}$. On the target subject, the modified cross-entropy loss is used to prevent gradient vanishing and exploding for adversarial learning \cite{goodfellow2014generative}. Then, we have
\begin{equation}
	\label{18}
	\mathcal{T}\big(\bb{f}(\phi(\bb{x})), h_{\bb{f'}_j}(\phi(\bb{x}))\big)=\log \Big[1- \Theta_{h_{\bb{f'}_j}}\big(\bb{f}(\phi(\bb{x}))\big)\Big].
\end{equation}

\tc{Therefore, the optimization problem in Eq. \eqref{14} can be written as the following two parts of optimization problem:}
\begin{equation}
	\label{19}
	\begin{aligned}
		\max_{\bb{f'}_j} \quad \mathcal{P}_{\widehat{P}_j, \widehat{Q}}^\mu =& \mathbb{E}_{\bb{x}\sim \widehat{Q}}\Big\{\log \Big[1- \Theta_{h_{\bb{f'}_j}}\big(\bb{f}(\phi(\bb{x}))\big)\Big]\Big\}\\
		&+\gamma\mathbb{E}_{\bb{x}\sim \widehat{P}_j} \Big\{\log \Big[ \Theta_{h_{\bb{f'}_j}}\big(\bb{f}(\phi(\bb{x}))\big)\Big]\Big\},
	\end{aligned}
\end{equation}
\begin{equation}
	\label{20}
		\min_{\phi,\bb{f}} \quad \mathcal{L}_{\phi,\bb{f}}=\sum_{j=1}^N \alpha_j\Big\{\mathcal{L}_{{\widehat{P}}_j}^{\mu}\left(\bb{f}\right)+\mathcal{P}_{\widehat{P}_j, \widehat{Q}}^\mu\left({\bb{f}},\bb{f'}_j\right)\Big\},\quad
\end{equation}
where $\mathcal{L}_{{\widehat{P}}_j}^{\mu}\left(\bb{f}\right)=\mathbb{E}_{(\bb{x},y)\sim \widehat{P}_j} \left\{-\log\big[ \Theta_y\big(\bb{f}(\phi(\bb{x}))\big)\big]\right\}$. 
During the iterative optimization process, the auxiliary function $\bb{f'}_j$ is firstly updated, and then scoring function $\bb{f}$ and feature extractor $\phi$ are updated.
\vspace{6pt}

\begin{theorem}
	\label{the_4}
	For optimization problem defined in Eq. \eqref{19}, fixing the classifier $\bb{f}$, then the estimated margin disparity discrepancy between source distribution $P$ and target distribution $Q$ is equivalent to
 $$
 \gamma \log\gamma-(1+\gamma)\log(1+\gamma)+ \gamma KL(P||Z)+KL(Q||Z),
 $$
 where $KL$ denotes the Kullback–Leibler divergence and $Z=(\gamma P+Q)/(\gamma+1)$ is the mixed distribution of $P$ and $Q$. 
\end{theorem}
\vspace{6pt}
\begin{remark}
    The proof of Theorem \ref{the_4} can be found in Appendix \ref{proof_the_4}. Since the Kullback–Leibler divergence between two distributions is always non-negative and zero only when they are equal, $\gamma \log\gamma-(1+\gamma)\log(1+\gamma)$ is the global minimum of MDD and that the only solution is $P=Q$. This indicates the different choices of $\gamma$ do not result in the mismatch between source distribution $P$ and target distribution $Q$. When the $\gamma$ is fixed, the estimated MDD can effectively reflect the gap between $P$ and $Q$.
\end{remark}

\subsection{Adversarial Learning Between Generator and Classifier}
In practical implementation, feature extractor $\phi(\cdot)$ is approximated by a convolution neural network (CNN). The scoring function $\bb{f}$ and auxiliary function $\bb{f'}_j$ $(j=1,...,N)$ are approximated by a fully connected neural network (NN). The $\bb{f}$ and $\bb{f'}_j$ have the same network frame. \tc{In the following part, we denote $\phi(\cdot)$ as the feature generator, $\bb{f}$ as the classifier, and $\bb{f'}_j$ as the auxiliary classifier.}

\tx{Due to the high real-time requirements of human intention recognition in practical applications, a lightweight neural network is used here.} Zhang \e \cite{zhang2023ensemble} pointed out that lightweight NN may have a better generalization ability, but may not fit the training dataset as accurately as a large NN. The ensemble method is commonly used to solve the above problem since it can enhance weak learners to make precise predictions \cite{zhou2012ensemble}. 

To trade off generalization ability and fitting ability, one tiny feature generator $\phi(\cdot)$ and $K$ tiny classifiers $\bb{F} = \{\bb{f}^{(i)}\}_{i=1}^{K}$ are employed here. Then, the classification results can be represented as
\begin{equation}
	\label{21}
	y = \bb{M}\left\{h_{\bb{f}^{(i)}}\big(\phi(\bb{x})\big)\right\}_{i=1}^K,
\end{equation}
where $\bb{M}$ denotes finding the mode from set. To further enhance the generalization of classifiers in ensemble learning, we firstly introduce classifier discrepancy $\mathcal{D}_{Q}^K({\bb{f}})$, which measures the classification difference between $K$ classifiers on target dataset $Q$, denoted as
\begin{equation}
	\label{22}
	\mathcal{D}_{Q}^K({\bb{f}})= \mathbb{E}_{\bb{x}\sim Q}\left\|\bb{F}\big(\phi(\bb{x})\big)-\text{mean}(\bb{F}) \right\|_1,
\end{equation}
where $\bb{F}(\cdot)=\{\bb{f}^{(1)}(\cdot),...,\bb{f}^{(K)}(\cdot)\}$ and $\|\cdot\|_1$ is $L_1$ norm. \tx{Then, motivated by the generative adversarial network (GAN) \cite{goodfellow2014generative}, adversarial learning is introduced to further update feature generator $\phi$ and classifiers $\{\bb{f}^{(i)}\}_{i=1}^K$ to improve the generalization ability, described as}
\begin{equation}
	\label{25}
	\mathop{\max}\limits_{\bb{f}} \mathop{\min}\limits_{\phi} \text{ } \mathcal{D}_{Q}^K({\bb{f}}).
\end{equation}

Since employing a classifier that is trained on the source domains with labels and the target domain without labels to predict target labels has high uncertainty, we maximize classifier discrepancy to improve ``exploration ability'' of ensemble classifiers on the target domain while ensuring the fitting ability on source domains. Therefore, the learning goal of ensemble classifiers in this stage can be denoted as
\begin{equation}
	\label{23}
\min_{\bb{f}} \quad \mathcal{L}_{\bb{f}}=\mathcal{L}_{\widehat{P}}^\mu(\bb{f})-\eta \mathcal{D}_{\widehat{Q}}^K({\bb{f}}),
\end{equation}
where $\eta$ trades off fitting ability and generalization ability. The goal of feature generator $\phi$ is to maximize ``universal ability'' of extracted features through $\phi$. That is minimizing classifier discrepancy when inputting extracted features into ensemble classifiers, that is
\begin{equation}
	\label{24}
	\min_{\phi} \quad \mathcal{L}_{\phi} =\mathcal{D}_{\widehat{Q}}^K({\bb{f}}).
\end{equation}

\tx{In adversarial learning, the ensemble classifiers are updated firstly fixing the feature generator, and then the feature generator is trained.} The fitting ability and generalization ability can achieve a balance through constant learning.

\begin{figure*}
	\centering
	\includegraphics[width=1\textwidth]{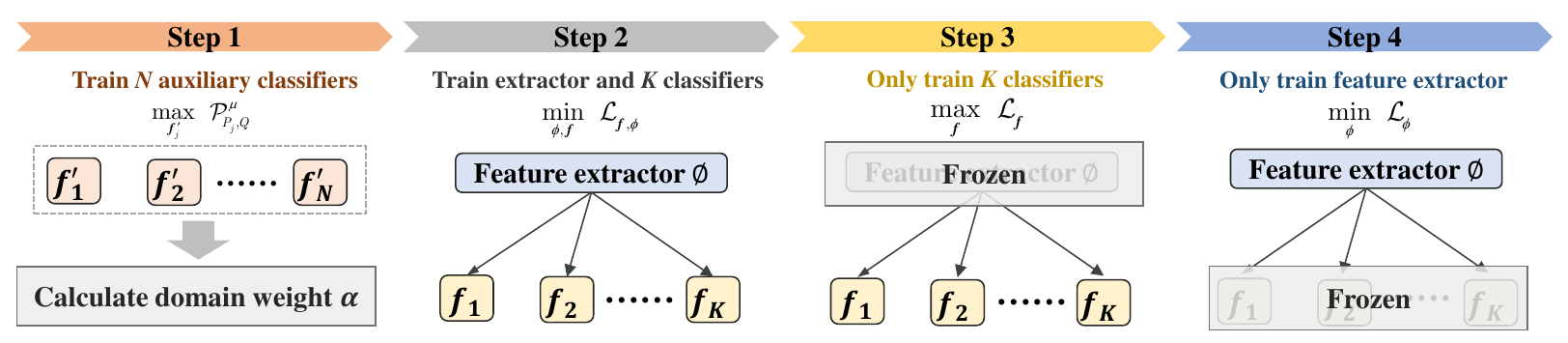}
	\caption{\tc{The training steps of the proposed algorithm WMDD. $\bb{f}'$, $\phi$ and $\bb{f}$ represent auxiliary classifiers, feature generator and classifier, respectively. Steps 1 to 4 are repeated in sequence until the training is completed.}}
	\label{figs1}
\end{figure*}

\begin{figure}[!t]
	\removelatexerror
\begin{algorithm}[H]
	 \label{alg1}
	\caption{WMDD}
	\begin{algorithmic}[1] 
		\STATE \textbf{Input}: $N$ source datasets $\{\mathcal{D}_{sj}\}_{j=1}^N$, one target dataset $\mathcal{D}_t$, feature generator $\phi$, $K$ classifiers $\{\bb{f}^{(i)}\}_{i=1}^K$ and $N$ auxiliary classifiers $\{\bb{f'}_{j}\}_{j=1}^N$.
		\STATE \textbf{Initialization:} Randomly initialize all networks.
		\FOR {$t=1,2,\cdots, n_{\text{iter}} $}
		\STATE Draw batch samples from target dataset $\mathcal{D}_{t}$.
		\FOR {$j=1,2,\cdots, N$}
		\STATE Draw batch samples from $j$-th source dataset $\mathcal{D}_{sj}$.
		\STATE Update auxiliary classifier $\bb{f'}_{j}$ according to Eq. \eqref{19}.
		\ENDFOR
		\STATE Compute source weight $\bb{\alpha}$ according to Eq. \eqref{15}.
		\STATE Update feature generator $\phi$ and classifiers $\{\bb{f}^{(i)}\}_{i=1}^K$ according to Eq. \eqref{20}.
		\STATE Update classifiers $\{\bb{f}^{(i)}\}_{i=1}^K$ according to Eq. \eqref{23}.
		\STATE Update feature generator $\phi$ according to Eq. \eqref{24}.
		\ENDFOR
	\end{algorithmic}
\end{algorithm}
\end{figure}

\subsection{The Details of Algorithm}
According to the above analysis, we propose a new \textbf{W}eighted multi-source UDA algorithm based on \textbf{M}argin \textbf{D}isparity \textbf{D}iscrepancy (WMDD). \tc{Fig. \ref{figs1} shows the detailed training steps of algorithm WMDD. It contains 4 steps. First, $N$ auxiliary classifiers are trained to estimate MDD, and then the source domain weights $\alpha$ are calculated. Second, the feature generator and $K$ classifiers are optimized based on minimizing generalization error. The last two steps are adversary learning, and the feature generator and $K$ classifiers are updated separately. Steps 1 to 4 are repeated in sequence until the training is completed. The detailed illustrations are given below.}

\begin{itemize}
    \item \textbf{Step 1:} The $j$-th auxiliary classifier $\bb{f'}_{j}$ is optimized to maximize the discrepancy defined in Eq. \eqref{19}. Then the source domain weight $\bb{\alpha}$ is obtained by the estimated MDD according to Eq. \eqref{15}.
    \item \textbf{Step 2:} The feature generator $\phi$ and classifiers $\{\bb{f}^{(i)}\}_{i=1}^K$ are optimized to minimize generalization error (target source) $\mathcal{L}_{\bb{f},\phi}$ in Eq. \eqref{20}.
    \item \textbf{Step 3:} Fixing feature generator $\phi$, then classifiers $\{\bb{f}^{(i)}\}_{i=1}^K$ are updated by maximizing classifier discrepancy through Eq. \eqref{23}.
    \item \textbf{Step 4:} Fixing classifiers $\{\bb{f}^{(i)}\}_{i=1}^K$, then feature generator $\phi$ is updated by minimizing classifier discrepancy $\mathcal{L}_{\phi}$ according to Eq. \eqref{24}.
\end{itemize}

\tc{The state of the data flow is as follows: Initially, multimodal data undergoes feature extraction through a feature generator network. Subsequently, the extracted feature information diverges into two pathways. The first pathway processes the data through an auxiliary classifier network to determine the source domain weights, while the second pathway feeds the data into a classifier network to obtain prediction outcomes. The visualization of specific data flow can be found in Fig. \ref{fig1}. The training procedures for each component of the network are detailed in Algorithm \ref{alg1}. The final WMDD method can use a feature generator and ensemble classifiers to classify the target subject intention accurately.}

	\section{Experiments}\label{sec:experiments}
This section gives the details of dataset and algorithm implementation, and focuses on answering the following questions through experiments: 

\begin{itemize}
	\item $\mathbf{Q_1}$: How does WMDD compare with current unsupervised domain adaptation methods for human motion recognition in standard benchmarks?
	\item $\mathbf{Q_2}$: How much does WMDD performance improve with source domain weight $\bb{\alpha}$ and adversarial learning between feature generator and classifier?
	\item $\mathbf{Q_3}$: How do hyperparameters margin coefficient $\gamma$, the number of classifiers $K$ and trade-off coefficient $\eta$ affect the performance of the WMDD algorithm?
\end{itemize}

\subsection{Experimental Setup}

\begin{table}
	\centering
	\normalsize
	\caption{Base hyperparameters of WMDD.}
	\label{tab_1}
	\renewcommand{\arraystretch}{0.9}
	\tabcolsep=0.7cm
		\begin{tabular}{ll}
			\toprule[1pt]
			\midrule
			\textbf{Parameter} & \textbf{Value} \\
			\midrule
			Optimizer & Adam\\	
			Batch size & $256$ \\
			Learning rate & $2\times 10^{-4}$\\
			Number of iterations $n_{\text{iter}}$& $400$\\
			Number of ensemble classifiers $K$ & $5$\\
            Number of auxiliary classifiers $N$ & $9/7$\\
			Trade-off coefficient $\eta$ & $5$\\
			Margin coefficient $\gamma$ & $0.1$\\
			\midrule
			\bottomrule[1pt]
		\end{tabular}
\end{table}

\textbf{Datasets}: In this paper, two public datasets are utilized to answer the above questions. One is the encyclopedia of able-bodied bilateral lower limb loco-motor signals (ENABL3S) collected by Northwestern University \cite{hu2018benchmark}. The other is the daily and sports activities data set (DSADS) collected by Bilken University \cite{barshan2014recognizing}. 

For the ENABL3S dataset \cite{hu2018benchmark}, ten subjects are invited to walk on several terrains and switch locomotion modes between sitting, standing, level ground walking, stair ascent, stair descent, ramp ascent, and ramp descent. It contains 7 class motion intentions in total, that is the above 7 activities. Each subject is asked to repeat walking on a circuit ten times. ENABL3S contains filtered EMG, IMU, and joint angle signals, which are segmented by a 300 ms wide sliding window, and the segmented signals are used to recognize human motion intention \cite{zhang2023ensemble}.

For the DSADS dataset \cite{barshan2014recognizing}, eight subjects are asked to perform 19 activities, including sitting, standing, running, riding a bike, jumping, and playing basketball. The DSADS contains five 9-axis IMU signals and the captured signals are segmented by 5 wide sliding window segments. Since there is no transition between different activities, the DASDS dataset is only utilized to classify the human motion modes \cite{zhang2020unsupervised1}.

\textbf{Experimental Details}: In MUDA experiments, a target subject (domain) is selected first and then the remaining subjects are used as source subjects. ENABL3S and DSADS datasets contain 22,000 and 9000 signal segments, respectively. The data from each subject are randomly shuffled and divided into a training set (70\%) and a test set (30\%). \tx{Table \ref{tab_1} gives the base hyper-parameters of the WMDD method. The learning rate is set as $2\times 10^{-4}$, the optimizer is Adam, the number of iterations is $400$ and the batch size is set as $256$. The margin coefficient $\gamma$ is set to $0.1$. The number of auxiliary classifiers $N$ is equal to the number of source subjects and set as $9$ and $7$ for ENABL3S and DSADS datasets, respectively.} Note that WMDD sets the same parameters for ENABL3S and DSADS datasets, which is different from previous methods that need setting different parameters for different datasets. This is also the advantage of the proposed method WMDD.

The network framework information is as follows. \tx{The lightweight network is composed of one feature generator, $K$ classifiers, and $N$ auxiliary classifiers, where the feature generator contains 3 convolution layers and the classifier includes 3 fully connected layers. The number of network parameters is $2.21$ M and $1.62$ M for ENABL3S and DSADS datasets, respectively.} The performance of WMDD has a close connection with domain weight $\bb{\alpha}$, number of ensemble classifiers $K$, and margin coefficient $\gamma$. The related experiments can be found in part C in this section. The code for WMDD is available at \href{https://github.com/xiaoyinliu0714/WMDD}{github.com/xiaoyinliu0714/WMDD}. The code is run under an Intel(R) Xeon(R) Gold 6348 CPU @ 2.60GHz and an NVIDIA GeForce RTX 4090 GPU. 

\begin{figure}
	\centering
	\includegraphics[width=0.46\textwidth]{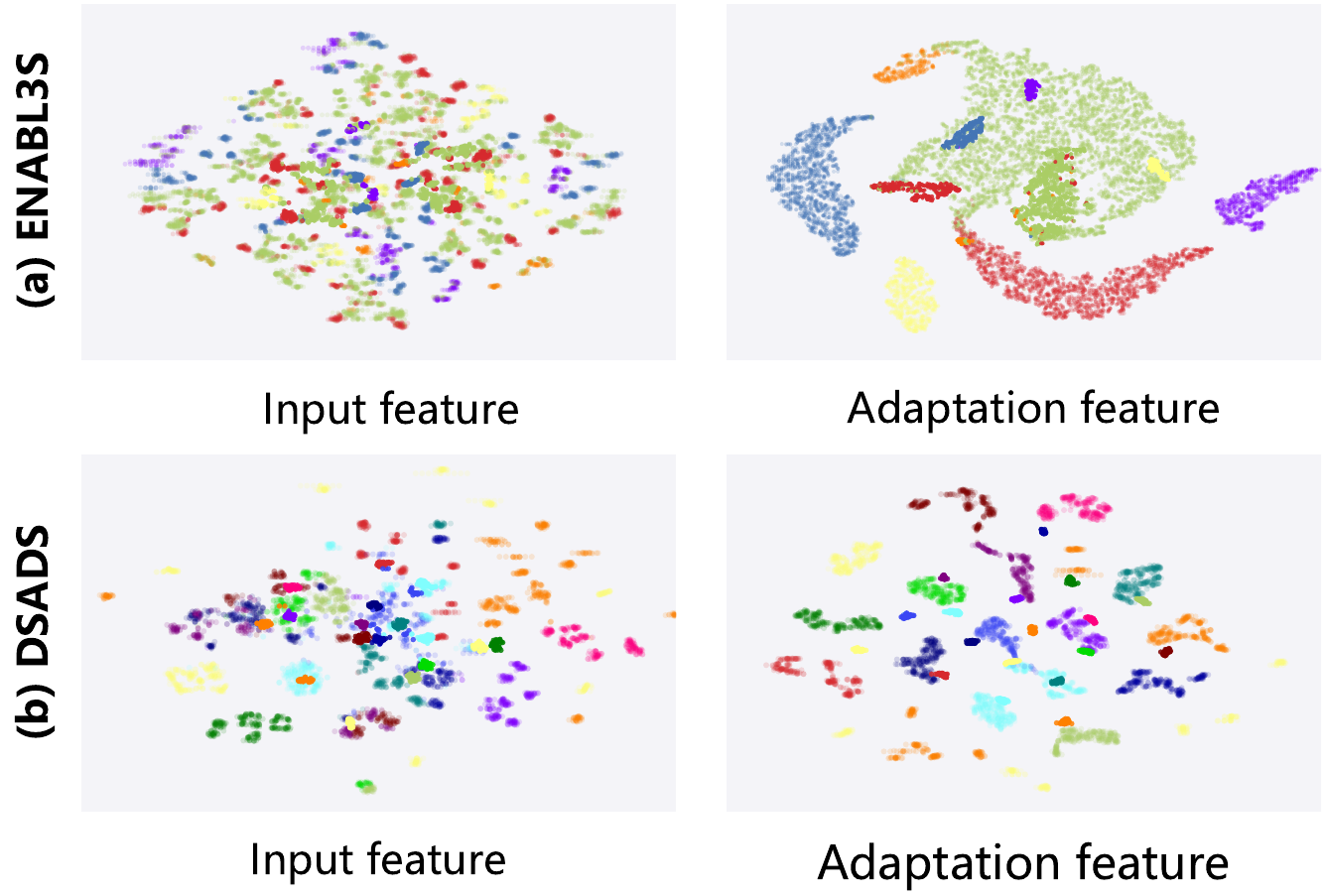}
	\caption{The visualization of t-SNE projection of non-adapted input features and the hidden features adapted by the feature generator. All features are extracted from the training set for ENABL3S (a) and DSADS (b). The different color points represent different classes. The dark-color and light-color points denote the target and source features, respectively}
	\label{fig3}
\end{figure}

\subsection{Feature Alignment and Classification Results}
Since several subjects are in one dataset, each subject is selected as a target subject sequentially to build training and testing datasets. Therefore, the average classification accuracy of each target subject is used as the final evaluation indicator.

\textbf{The Results of Feature Alignment}: The dimensional reduction method t-SNE \cite{van2008visualizing}, which can keep the clustering of high dimensional space, is used to visualize the results of feature extraction following previous work \cite{zhang2023ensemble}. Fig. \ref{fig3} shows the t-SNE projection of the non-adapted input features and the hidden features adapted by WMDD. It can be found that the input features are not clustered, and the source distribution is not aligned with the target distribution. After feature alignment, the features of the same class almost cluster together for two datasets. For the DSADS dataset, source and target features that belong to the same class are almost in the same region. \tx{However, for the ENABL3S dataset, some source and target features that belong to the same class aren't in the same region, which causes the classification accuracy of the ENABL3S dataset to be worse than that of the DSADS dataset. The reason is that about 30\% of data for ENABL3S are transition activities instead of steady activities, where the activity is defined as a transition activity if the locomotion mode of the current activity is different from that of the last activity.} 


\begin{table*}
	\small
	\renewcommand{\arraystretch}{1.2}
	\caption{The detailed results for different target subjects and corresponding training and testing times for ENABL3S and DSADS datasets. Std indicates the standard deviation. S1 represents the subject 1, which means the first subject is the target subject and other subjects are source subjects.}
        \label{tab_result}
	\centering
	\begin{tabular}{c|cccccccccc|cc|cc}
		\toprule[1pt]
		\midrule
\textbf{Dataset}&S$1$&S$2$&S$3$&S$4$&S$5$&S$6$&S$7$&S$8$&S$9$&S$10$&\textbf{Means}&\textbf{Std}&\textbf{Training}&\textbf{Testing}\\
  
\midrule
ENABL3S&$97.8$&$95.3$&$94.9$&$91.5$&$94.6$&$94.2$&$91.3$&$95.4$&$96.7$&$94.5$&$\mathbf{94.6}$&$2.0$&$15.4 \text{ min}$&$0.54 \text{ ms}$\\
  DSADS&$99.1$&$99.7$&$98.8$&$99.7$&$99.4$&$99.4$&$98.2$&$97.7$&--&--&$\mathbf{99.0}$&$0.7$&$11.3\text{ min}$&$0.49 \text{ ms}$\\
  \midrule
		\bottomrule[1pt]
	\end{tabular}
\end{table*}

\begin{table}
	\centering
	\caption{The mean accuracy of classification for the target subject of ENABL3S and DSADS using different domain adaptation methods. Std indicates the standard deviation.}
	\label{tab_2}
	\renewcommand{\arraystretch}{1.1}
	\resizebox{1\linewidth}{!}{
		\begin{tabular}{ccccc}
			\toprule[1pt]
			\midrule
			\multirow{2}{*}{\textbf{Method}}&\multicolumn{2}{c}{\textbf{ENABL3S} }& \multicolumn{2}{c}{\textbf{DSADS} }\\
			\cmidrule{2-5}
			&\textbf{Mean(\%)}&\textbf{Std(\%)}&\textbf{Mean(\%)}&\textbf{Std(\%)}\\
			\midrule
			DANN \cite{ganin2016domain}&$88.5$&$4.8$&$91.1$&$5.2$\\
			MMD \cite{long2017learning}&$92.7$&$2.2$&$95.4$&$3.3$\\
			MCD \cite{saito2018maximum}&$93.9$&$1.8$&$95.3$&$4.5$\\
			DFA \cite{wang2021discriminative}&$91.8$&$3.0$&$92.5$&$3.3$\\
			GFA \cite{zhang2022gaussian}&$93.7$&$1.7$&$96.9$&$3.3$\\
			EDHKD \cite{zhang2023ensemble}&$94.4$&$1.7$&$97.4$&$4.9$\\
			\textbf{WMDD} (Ours)&$\mathbf{94.6}$&$2.0$&$\mathbf{99.0}$&$0.7$\\
			\midrule
			\bottomrule[1pt]
		\end{tabular}
	}
\end{table}

\textbf{{The Classification Results on ENABL3S and DSADS}}: To answer $\mathbf{Q_1}$, we compare different domain adaptation methods in ENABL3S and DSADS datasets, including DANN \cite{ganin2016domain}, MMD \cite{long2017learning}, MCD \cite{saito2018maximum}, DFA \cite{wang2021discriminative}, GFA \cite{zhang2022gaussian} and EDHKD \cite{zhang2023ensemble}, where EDHKD is the state-of-the-art method for HMI using domain adaptation method. The classification results on ENABL3S and DSADS of these methods come from article \cite{zhang2023ensemble}. \tc{Table \ref{tab_result} gives the detailed results for different target subjects on ENABL3S and DSADS datasets, and shows the related training and testing times of lightweight network. It indicates that the performance of WMDD varies slightly across different target subjects, but it can achieve a relatively high average performance. Additionally, the training process of the WMDD requires $15$ and $11$ minutes on ENABL3S and DSADS datasets, indicating low computational cost. Moreover, the testing time of for a single sample is about $0.5$ ms, which can guarantee the real-time of the testing.}

Table \ref{tab_2} compares the mean accuracy of classification on the target subject for different domain adaptation methods. It shows that WMDD achieves the accuracy of $94.6\% \pm 2.0\%$ and $99.0\% \pm 0.7\%$ for target subject in ENABL3S and DSADS datasets respectively, which are $0.2\%$ and $1.6\%$ higher than the state-of-the-art HMI result using domain adaptation methods. This verifies that WMDD can achieve a better performance than previous methods in both ENABL3S and DSADS datasets. \tx{However, the improvement in ENABL3S is not significant despite that the samples of ENABL3S are larger than DSADS. The reason lies in the effects of transition activities. The results of classification are consistent with that of feature alignment. In addition, we give the final source domain weight $\bb{\alpha}$ when the first subject is the target subject. The weights are $0.1108$, $0.1106$, $0.1108$, $0.1106$, $0.1110$, $0.1118$, $0.1121$, $0.1114$, $0.1108$ for ENABL3S dataset and $0.1438$, $0.1431$, $0.1417$, $0.1454$, $0.1441$, $0.1396$, $0.1424$ for DSADS dataset.}

\tx{The superior performance of our method compared to EDHKD can be attributed to two key factors: 1) EDHKD does not adequately account for the variations among individual source subjects, which may negatively impact classification accuracy. In contrast, our approach explicitly considers these inter-source differences, leading to improved classification outcomes. 2) We extend the margin disparity discrepancy (MDD) to the context of multi-source unsupervised domain adaptation. MDD incorporates more comprehensive generalization bound information, where the generalization bound is affected by the margin parameter $\mu$. By selecting an appropriate $\mu$, the classification error on the target subject can be minimized, thereby enhancing performance on the target domain.}

\subsection{Ablation Experiments} \label{sec:1}
\textbf{Effects of Weight $\bb{\alpha}$ and Adversarial Learning}: To answer $\mathbf{Q_2}$, the ablation experiments for two parts are designed (see Table \ref{tab_4}). The weight $\bb{\alpha}$ reflects the discrepancy between different source domains and target domain. The adversarial learning is used to improve the classifier's performance. \tx{From this table, the average performance in ENABL3S and DSADS is improved by $1.07\%$ and $1.02\%$ respectively compared with method No-W. The average performance is improved by $1.72\%$ and $7.25\%$ respectively compared with method No-A, and the maximum performance improvement reaches $5.91\%$ on one of the target subject. Therefore, the usage of domain weight and adversarial learning can both improve the performance of the classifier, and the effect of adversarial learning is more significant.}

\begin{figure*}
	\centering
	\includegraphics[width=1\textwidth]{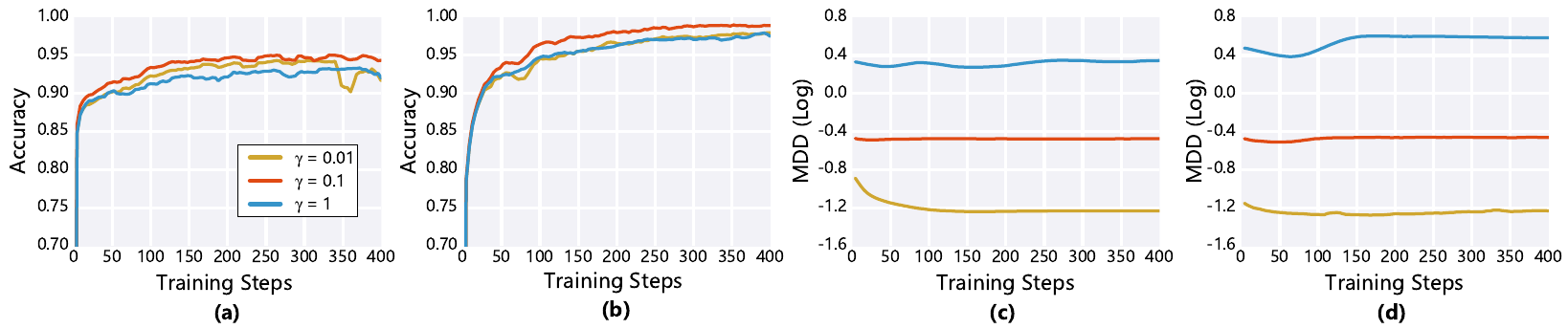}
	\caption{The comparison results of training curve under different margin $\mu$ on ENABL3S (a) and DSADS (b) datasets. The comparison results of the average margin disparity discrepancy between source and target distribution under different $\mu$ on ENABL3S (c) and DSADS (d) datasets. For the convenience of comparison, the ordinate values of sub-figure (c) and (d) are $\log{\text{MDD}}$.}
	\label{fig5}
\end{figure*}

\begin{table*}
	\small
	\renewcommand{\arraystretch}{1}
	\caption{The relationship between the classification accuracy and the number of classifiers $K$, and the relationship between classification accuracy and trade-off coefficient $\eta$ on ENABL3S and DSADS datasets.000}
        \label{ablation}
	\centering
	\begin{tabular}{cccccccccccccc}
		\toprule[1pt]
		\midrule
\textbf{Number of classifiers} $K$ &$1$&$3$&$5$&$7$&$9$&$11$&$13$&$15$&$17$&$19$&$21$&$23$&$25$\\ 
  \midrule
\textbf{ENABL3S}&$93.3$&$91.8$&$\mathbf{97.8}$&$96.4$&$96.5$&$96.5$&$95.4$&$95.9$&$96.2$&$95.6$&$95.8$&$96.1$&$96.4$\\
\textbf{DSADS}&$96.8$&$98.0$&$\mathbf{99.1}$&$98.2$&$97.7$&$98.0$&$98.5$&$98.0$&$\mathbf{99.1}$&$97.4$&$98.2$&$98.2$&$97.4$\\
  \midrule  \midrule
\textbf{Trade-off coefficient} $\eta$  &$1$&$2$&$3$&$4$&$5$&$7$&$8$&$9$&$11$&$12$&$13$&$14$&$15$\\
  \midrule
\textbf{ENABL3S}&$96.4$&$93.6$&$96.7$&$\mathbf{97.6}$&$\mathbf{97.8}$&$97.0$&$96.5$&$97.3$&$96.1$&$96.4$&$96.4$&$94.7$&$94.0$\\
\textbf{DSADS}&$94.4$&$98.5$&$97.4$&$99.1$&$\mathbf{99.4}$&$99.1$&$99.1$&$99.1$&$97.7$&$98.5$&$96.2$&$98.2$&$92.1$\\
  \midrule
		\bottomrule[1pt]
	\end{tabular}
\end{table*}

\begin{table}
	\centering
	\caption{The ablation experiments for domain weight and adversarial learning. No-W and No-A represent that the domain weight method and adversarial learning are not employed, respectively. No-W-A represents both methods are not used. }
	\label{tab_4}
	\renewcommand{\arraystretch}{1.1}
	\resizebox{1\linewidth}{!}{
		\begin{tabular}{ccccc}
			\toprule[1pt]
			\midrule
			Name &Weight $\bb{\alpha}$&Adversary &ENABL3S&DSADS\\
			\midrule
			No-W-A&&&$93.1\pm2.2$&$92.7\pm4.2$\\
			No-W&&\Checkmark&$93.6\pm2.3$&$98.0\pm2.1$\\
			No-A&\Checkmark&&$93.0\pm2.6$&$92.3\pm4.4$\\
			Ours&\Checkmark&\Checkmark&$\mathbf{94.6\pm2.0}$&$\mathbf{99.0\pm0.7}$\\
			\midrule
			\bottomrule[1pt]
		\end{tabular}
	}
\end{table}


\textbf{Effect of Margin Coefficient $\gamma$}: The margin coefficient $\gamma$ determines the margin disparity discrepancy between source and target distributions and influences the classification accuracy on the target subject, that is generalization error. \tc{Theorem \ref{the_4} gives the equivalent form of MDD. It reflects the minimum of MDD is $\gamma \log\gamma-(1+\gamma)\log(1+\gamma)$ when the source distribution equals the target distribution. The margin coefficient $\gamma$ only influences the value of MDD and doesn't affect the gap between source and target distributions.} 

To answer $\mathbf{Q_3}$, we give the training curve and MDD under margin coefficient $\gamma = 0.01, 0.1, 1$ on ENABL3S and DSADS datasets. Fig. \ref{fig5} shows that the method WMDD achieves better classification accuracy under margin coefficient $\gamma=0.1$ for ENABL3S and DSADS datasets. The recognition accuracy are $91.6\%$, $94.6\%$, $92.0\%$ under $\gamma = 0.01$, $0.1$, $1$ respectively for ENABL3S dataset, and are $97.9\%$, $99.0\%$, $97.5\%$ for DSADS dataset. \tc{The average values of the estimated MDD are $0.06$, $0.34$, $2.20$ under $\gamma = 0.01$, $0.1$, $1$ respectively for ENABL3S dataset, and are $0.06$, $0.35$, $3.84$ for DSADS dataset. This result indicates that the average MDD between each source subject and target subject increases as margin coefficient $\gamma$ increases. However, for too large $\gamma$, the recognition accuracy cannot be improved. This shows that the margin $\mu$ in Theorem \ref{the_2} cannot be set too large.}


\textbf{Effects of the Number of Classifiers $K$ and Trade-off Coefficient $\eta$}: To answer $\mathbf{Q_3}$, the more experiments for the number of classifiers $K$ and trade-off coefficient $\eta$ are conducted below. \tc{The more number of classifiers $K$ often consume more computation cost. It is a necessity to choose an appropriate number of classifiers $K$ to achieve a balance between the classification performance and computation cost. The coefficient $\eta$ is used to trade off the fitting ability and generalization ability of the network. The performance of WMDD is determined by this hyperparameter.}

\tc{Table \ref{ablation} shows the relationship between the classification accuracy and the number of classifiers $K$, and the relationship between classification accuracy and trade-off coefficient $\eta$ on ENABL3S and DSADS datasets. 
This suggests that the increase in the number of classifiers $K$ does not necessarily lead to improved classification performance. Compared with the accuracy of $5$ classifiers, the accuracy of $25$ classifiers decreases by $1.56\%$ and $5.61\%$ for ENABL3S and DSADS datasets respectively. For lightweight networks, classification performance can be effectively enhanced by selecting an optimal number of classifiers. Moreover, the results indicate that the relatively optimal parameters for $K$ and $\eta$ are both $5$ on ENABL3S and DSADS datasets. If coefficient $\eta$ is too small, the network's fitting ability will improve, but its generalization capability will decline. Conversely, if coefficient is too large, the generalization ability will increase, while its fitting ability will diminish. Therefore, too large or too small coefficient $\eta$ will decrease the performance of the algorithm.}

\section{Discussion}\label{Discussion}
This paper aims to extend MDD theory to multi-source UDA, and propose a novel multi-source UDA algorithm for HMI recognition based on the derived theory. \tx{In the below part, we give further discussions for multi-source UDA theory and algorithm, and discuss more related works about unsupervised domain adaption.}

\subsection{The Discussion for Theory}
The developed theory in this paper is based on \cite{ben2010theory,zhang2019bridging}. It can effectively answer the challenges listed in section \ref{sec:introduction}. \textit{Challenge 1:} The margin disparity discrepancy is used to measure the gap between source and target domains. The MDD has below features. 1) The MDD contains more generalization bound information. Theorem \ref{the_2} shows the generalization bound is influenced by the margin $\mu$. The better performance on the the target subject can be achieved through choosing
appropriate margin $\mu$. 2) The MDD can accurately measure the gap between different distributions. Theorem \ref{the_4} shows the estimated MDD reflects the gap through KL divergence, which avoids the error brought by estimating KL divergence. 

\textit{Challenge 2:} The dataset is collected from multiple source domains that might be different not only from the target domain but also from each other. The source domain weight determined by MDD is incorporated into Theorem \ref{the_3} to measure the difference between source domains, where the source domain with the small gap to the target domain should have a large weight. The weight is adjusted adaptively by MDD according to Eq. \eqref{15}. Since the training data is randomly selected, the drawn data hardly fully describe the entire distribution characteristic. Adjusting weight dynamically according to the drawn data is beneficial to improving performance on the target domain. Table \ref{tab_4} also verifies this conclusion.
\vspace{0.15cm}

\subsection{The Discussion for Algorithm} 
Since the margin loss easily leads to gradient vanishing, the cross-entropy loss is employed to replace the margin loss in this paper. The $\gamma$ in Eq. \eqref{19} reflects the margin $\mu$ in MDD. Theorem \ref{the_4} shows that the result of using a surrogate function can better measure the discrepancy between different distributions and maintain the main characteristics of MDD through $\gamma$. \tx{The algorithm is essentially a two-stage game problem. The first stage is the game problem between the auxiliary classifiers and classifiers, and the second stage is the game problem between the classifiers and feature generator. In the first stage, the auxiliary classifiers are first updated and the MDD between source and target domain is estimated. Then classifiers and feature generator are updated together through minimizing the generalization error. In the second stage, the feature generator and classifiers are the adversarial parties. The generalization ability (performance on target domain) is improved through constant adversarial learning between feature generator and classifiers.}

\subsection{\tc{The Discussion for Unsupervised Domain Adaption}}
\tx{Domain shift is a fundamental challenge for recognition and prediction, where there is a gap between the training (source) domain and test (target) domain. Domain adaption method and meta learning technique have been utilized to deal with the above problem. Meta learning, such as MAML \cite{finn2018meta}, acquires knowledge from multiple tasks, enabling faster adaptation and generalization to new tasks. Domain adaption can be considered as part of meta learning when treating domain as a particular task. The most unsupervised domain adaption methods \cite{long2017learning,saito2018maximum}, including our method, aim to achieve feature alignment between source and target domain. However, the performance of these methods would be limited if data in source domain is insufficient \cite{vettoruzzo2024advances}. Some meta learning techniques \cite{yang2022few} can be combined with domain adaptation to solve the above problem. The goal of domain adaption and meta learning is same, that is pursuing the generality and adaptability in artificial intelligence.}

\tc{Moreover, unsupervised domain adaption technique has been widely applied in the cross-subject tasks. Some methods make improvements to the perception network framework itself, such as using multi-encoder autoencoder \cite{Gu2021}, to improve performance across subjects. Other methods improve the algorithm itself, such as modifying the loss function \cite{zhang2020unsupervised,deng2023mixture}, to minimize the error in the target subject and thus improve the recognition accuracy. Compared with the above cross-subject methods, the proposed method of this paper can fully integrate the differences between multiple subject data through adaptively adjusting the weight of different source subjects, and then the more effective information can be obtained for classification. Therefore, under the same amount of data, the recognition accuracy of the proposed method can be improved on the cross-subject tasks.}

\section{Conclusion}\label{Conclusion}
This paper developed a new theory for multi-source UDA based on margin disparity discrepancy and derived a novel generalization bound for multi-source UDA. Motivated by generalization bound, a novel weight-aware-based multi-source UDA algorithm (WMDD) was proposed for HMI recognition. The proposed method WMDD can improve classification accuracy of HMI recognition tasks through considering the difference between each source subject, where the source domain weight can be adjusted adaptively by the estimated discrepancy. WMDD can also guarantee the real-time of HMI recognition by utilizing a lightweight network. Extensive experiments confirmed that WMDD can achieve state-of-the-art accuracy on HMI recognition tasks. 

\tx{We expect the proposed theory and algorithm can provide reference and inspiration for other multi-source UDA theories and cross-subject application fields. However, the current domain adaptation methods require retraining the network for each new target subject to achieve better performance, which is computationally expensive. Future works can focus on exploring a domain-adaptive fine-tuning method that leverages unlabeled data from the target subject to fine-tune the base model, thereby enhancing performance.}


	\newpage
\clearpage
\setcounter{page}{1}

\begin{center}
    \Large\textbf{Supplementary Materials (Appendix)} \\
    \vspace{0.5cm}
\end{center}

\subsection{Proof of Theorem \ref{the_3}}\label{proof_3}
Before proving the Theorem \ref{the_3}, we firstly give the below Proposition similar to the previous work\cite{zhang2019bridging}.
\begin{proposition}\label{lem_1}
    Fix $\mu$. For any scoring function $\bb{f} \in \mathcal{F}$,
    \begin{equation*}
        \mathbb{W}_\mu\big[\omega_{\bb{f'}}\big(x,h_{\bb{f}}(x)\big)\big]\leq\mathbb{W}_\mu\big[\omega_{\bb{f}}\big(x,y\big)\big]+\mathbb{W}_\mu\big[\omega_{\bb{f'}}\big(x,y\big)\big],
    \end{equation*}
    where $\omega_{\bb{f}}(\bb{x},y)=[f_y(\bb{x})-\max_{y'\neq y}f_{y'}(\bb{x})]/2$ is the margin of $\bb{f}$ in sample $(x,y)$, and $\mathbb{W}_\mu(v)$ is the ramp loss that is defined in Eq. \eqref{4}.
\end{proposition}

\textit{\textbf{Proof.}} For any sample $(x,y)$, if $h_{\bb{f}}(x)\neq y$ or $h_{\bb{f'}}(x)\neq y$, $\omega_{\bb{f}}(x,y)$ or $\omega_{\bb{f}}(x,y)$ is small than zero, implying the right side of above
equation reach 1, and further deducing that the inequality always holds. Otherwise $h_{\bb{f}}(x)= y$ and $h_{\bb{f'}}(x)=y$, then we can derive
\begin{equation}
\begin{aligned}
            &\mathbb{W}_\mu\big[\omega_{\bb{f'}}\big(x,h_{\bb{f}}(x)\big)\big]\\
            \leq&\mathbb{W}_\mu\big[\omega_{\bb{f'}}\big(x,h_{\bb{f}}(x)\big)\big]+\mathbb{W}_\mu\big[\omega_{\bb{f}}\big(x,y\big)\big]\\
            \leq&\mathbb{W}_\mu\big[\omega_{\bb{f'}}\big(x,y\big)\big]+\mathbb{W}_\mu\big[\omega_{\bb{f}}\big(x,y\big)\big].
\end{aligned}
\end{equation}

This completes the proof of Proposition \ref{lem_1}. Next we give the proof for Theorem \ref{the_3}. The proof follows the previous works \cite{zhang2019bridging,zhang2020unsupervised}.
\vspace{10pt}

\textbf{{The Proof for Theorem \ref{the_3}}}:  Let be the ideal scoring function $\bb{f^{*}}$ which minimizes the combined margin loss,
\begin{equation}
    \label{a01}
    \bb{f^{*}}=\arg\min_{\bb{f}\in \mathcal{F}}\Big\{\mathcal{L}_{Q}^{\mu}\left(\bb{f}\right)+\mathcal{L}_{P}^{\mu}\left(\bb{f}\right)\Big\}.
\end{equation}

Then, for any $\bb{f}\in \mathcal{F}$, 
\begin{equation*}
    \label{a02}
    \begin{aligned}
        &\mathcal{L}_Q\left(h_{\bb{f}}\right)= \mathbb{E}_{Q} \big\{\mathbb{I}\left[h_{\bb{f}}(\bb{x})\ne y\right]\big\}\\
        \leq&\mathbb{E}_{Q} \big\{\mathbb{I}\left[h_{\bb{f}}(\bb{x})\ne h_{\bb{f^*}}(\bb{x})\right]\big\}+\mathbb{E}_{Q} \big\{\mathbb{I}\left[h_{\bb{f^*}}(\bb{x})\ne y\right]\big\}\\
        \leq&\mathbb{E}_{Q}\big\{\mathbb{W}_\mu\big[\omega_{\bb{f}^*}\big(x,h_{\bb{f}}(x)\big)\big]\big\}+  \underbrace{\mathbb{E}_{Q}\big\{\mathbb{W}_\mu\big[\omega_{\bb{f^*}}\big(x,y\big)\big]\big\}}_{:=\mathcal{L}_{Q}^{\mu}\left(\bb{f^*}\right)}\\
        &+\underbrace{\mathbb{E}_{P}\big\{\mathbb{W}_\mu\big[\omega_{\bb{f}}\big(x,y\big)\big]\big\}}_{:=\mathcal{L}_{P}^{\mu}\left(\bb{f}\right)}-\underbrace{\mathbb{E}_{P}\big\{\mathbb{W}_\mu\big[\omega_{\bb{f}}\big(x,y\big)\big]\big\}}_{\text{Proposition \ref{lem_1}}}\\
        \leq&\mathcal{L}_{P}^{\mu}\left(\bb{f}\right)+\mathcal{L}_{Q}^{\mu}\left(\bb{f^*}\right)+\underbrace{\mathbb{E}_{P}\big\{\mathbb{W}_\mu\big[\omega_{\bb{f^*}}\big(x,y\big)\big]\big\}}_{:=\mathcal{L}_{P}^{\mu}\left(\bb{f^*}\right)}\\
        &+\underbrace{\mathbb{E}_{Q}\big\{\mathbb{W}_\mu\big[\omega_{\bb{f^*}}\big(x,h_{\bb{f}}(x)\big)\big]\big\}-\mathbb{E}_{P}\big\{\mathbb{W}_\mu\big[\omega_{\bb{f^*}}\big(x,h_{\bb{f}}(x)\big)\big]\big\}}_{\text{Eq. \eqref{5}}}\\
        \leq 
        &\mathcal{L}_{P}^{\mu}\left(\bb{f}\right)+d^\mu_{\bb{f}}\left(P, Q\right)+\lambda,
    \end{aligned}
\end{equation*}
where the second inequality is the important property of margin loss for any $\mu$ and $\bb{f}$, and $\lambda=\mathcal{L}_{Q}^{\mu}\left(\bb{f^*}\right)+\mathcal{L}_{P}^{\mu}\left(\bb{f^*}\right)=\min_{\bb{f} \in \mathcal{F}}\{\mathcal{L}_{Q}^{\mu}\left(\bb{f}\right)+\mathcal{L}_{P}^{\mu}\left(\bb{f}\right)\}$. \tc{This completes the proof of Theorem \ref{the_3}.}

\subsection{Proof of Theorem \ref{the_1}}\label{proof_1}
\textbf{{The Proof for Theorem \ref{the_1}}}:
    Let $\widetilde{P}$ be the mixture distribution of the $N$ source domains, denoted as $\widetilde{P}=\sum_{j=1}^{N}\alpha_jP_j$, and $\mathcal{D}_{\tilde{s}}$ be the combined samples from $N$ source domains. We denote $\widetilde{P}$ and $Q$ as the source distribution and the target distribution in Theorem \ref{the_3}, respectively. Then, we have
\begin{equation}
    \label{a1}
    		\mathcal{L}_Q\left(h_{\bb{f}}\right) \leq \mathcal{L}_{\widetilde{P}}^{\mu}\left(\bb{f}\right)+d^\mu_{\bb{f}}\left(\widetilde{P}, Q\right)+\widetilde{\lambda}.
\end{equation}

On the one hand, for any $\bb{f}\in\mathcal{F}$, the following holds
\begin{equation}
    \label{a2}
\mathcal{L}_{\widetilde{P}}^{\mu}\left(\bb{f}\right)=\sum_{j=1}^{N}\alpha_j\mathcal{L}_{P_j}^{\mu}\left(\bb{f}\right),
\end{equation}
then $\widetilde{\lambda}=\min_{\bb{f}\in\mathcal{F}}\{\sum_{j=1}^{N}\alpha_j\mathcal{L}_{P_j}^{\mu}\left(\bb{f}\right)+\mathcal{L}_{Q}^{\mu}(\bb{f})\}$, we denote it as $\beta$ here. On the other hand, according to Eq. \eqref{5} the term $d^\mu_{\bb{f}}(\widetilde{P}, Q)$ can be upper bounded by
\begin{equation}
\begin{aligned}
       \label{a3}
d^\mu_{\bb{f}}\left(\widetilde{P}, Q\right)&=\sup _{\boldsymbol{f}^{\prime} \in \mathcal{F}}\Big\{\mathbb{E}_{Q}\big[ \mathbb{W}_\mu\big]-\mathbb{E}_{\widetilde{P}} \big[\mathbb{W}_\mu\big]\Big\}\\
&=\sup _{\boldsymbol{f}^{\prime} \in \mathcal{F}}\Big\{\mathbb{E}_{Q}\big[ \mathbb{W}_\mu\big]-\sum_{j=1}^{N}\alpha_j\mathbb{E}_{{P_j}} \big[\mathbb{W}_\mu\big]\Big\}\\
&=\sup _{\boldsymbol{f}^{\prime} \in \mathcal{F}}\Big\{\sum_{j=1}^{N}\alpha_j \Big(\mathbb{E}_{Q}\big[ \mathbb{W}_\mu\big]-\mathbb{E}_{{P_j}} \big[\mathbb{W}_\mu\big]\Big)\Big\}\\
&\leq \sum_{j=1}^{N}\alpha_j \sup _{\boldsymbol{f}^{\prime} \in \mathcal{F}} \Big\{\mathbb{E}_{Q}\big[ \mathbb{W}_\mu\big]-\mathbb{E}_{{P_j}} \big[\mathbb{W}_\mu\big]\Big\}\\
&= \sum_{j=1}^{N}\alpha_j d^\mu_{\bb{f}}\left({P_j}, Q\right),
\end{aligned}
\end{equation}
where the first inequality is by the sub-additivity of the sup function. Then bringing the Eqs. \eqref{a2} and \eqref{a3} into Eq. \eqref{a1}, we can get
\begin{equation}
		\label{a4}
			\mathcal{L}_Q\left(h_{\bb{f}}\right) \leq \sum_{j=1}^N \alpha_j\left(\mathcal{L}_{{P}_j}^{\mu}\left(\bb{f}\right)+d^\mu_{\bb{f}}\left(P_j, Q\right)\right)+\beta.
\end{equation}

This completes the proof of Theorem \ref{the_1}.

\subsection{Proof of Theorem \ref{the_2}}\label{proof_2}

\begin{lemma}
\label{lem_2}
    Let $\mathcal{G}$ be a family of functions mapping from $\mathcal{X} $ to $  [0,1]$ and $\widehat{\mathcal{D}}$ be empirical datasets sampled from an i.i.d. sample ${\mathcal{D}}$ of size $m$. Then, for any $\delta>0$, with probability at least $1-\delta$, the following holds for all $g\in \mathcal{G}$, 
    \begin{equation}
\Big|\mathbb{E}_{\mathcal{D}}\left[g\right]-\mathbb{E}_{\widehat{\mathcal{D}}}\left[g\right]\Big|\leq 2\widehat{\mathfrak{R}}_{{\widehat{\mathcal{D}}}}(\mathcal{G})+3\sqrt{\frac{\log (2 / \delta)}{2m}},
    \end{equation}
    where $\mathbb{E}_{\widehat{\mathcal{D}}}\left[g\right]=\frac{1}{m}\sum_{i=1}^{m}\left[g(x_i)\right]$ is the empirical form of $\mathbb{E}_{\mathcal{D}}\left[g\right]$. This proof can be found in Theorem 3.3 of work \cite{mohri2018foundations}.
\end{lemma}
\vspace{10pt}

\begin{lemma}\label{lem_3}
Let $\mathcal{G}$ be a family of functions, denoted as $\mathcal{G}=\{\max\{\bb{f}_1,\bb{f}_2,...,\bb{f}_{k}\}~|~\bb{f}_i \in \mathcal{F}, i \in \{1,2,...,k\}\}$. Then, for any sample ${\mathcal{D}}$ of size $m$, the following holds for all $g\in \mathcal{G}$,
    \begin{equation}
\widehat{\mathfrak{R}}_{{\widehat{\mathcal{D}}}}(\mathcal{G})\leq k\widehat{\mathfrak{R}}_{{\widehat{\mathcal{D}}}}(\mathcal{F}).
    \end{equation}
\end{lemma}

This Lemma is the modified version of Lemma C.6 in previous work \cite{zhang2019bridging}.
\vspace{10pt}

\begin{proposition}
    \label{pro_2}
    Let $\mathcal{G}$ be a family of margin loss functions defined in Eq. \eqref{3} and $\Omega_1(\mathcal{F})$ be a family of functions defined in Definition \ref{def_2}. Then, for any sample $\widehat{\mathcal{D}}=\{(\bb{x}_i,y_i)\}_{i=1}^{m}$, where $y_i\in \{1,...,\mathcal{C}\}$, the following relation holds between the empirical Rademacher complexities of $\mathcal{G}$ and $\Omega_1(\mathcal{F})$:
    \begin{equation}
    \widehat{\mathfrak{R}}_{{\widehat{\mathcal{D}}}}(\mathcal{G})\leq\frac{\mathcal{C}}{2\mu}\widehat{\mathfrak{R}}_{{\widehat{\mathcal{D}}}}(\Omega_1(\mathcal{F})).
    \end{equation}
\end{proposition}

\textit{\textbf{Proof.}} According to Definition \ref{def_1}, the empirical Rademacher complexity of $\mathcal{G}$ can be written as:
\begin{equation}
\begin{aligned}
    \label{a20}
		\widehat{\Re}_{\widehat{\mathcal{D}}}(\mathcal{G})=& \mathbb{E}_{\bb{\sigma}} \bigg[\sup _{\bb{f} \in \mathcal{F}} \frac{1}{m}\sum_{i=1}^m \sigma_i \mathbb{W}_\mu\big[\omega_{\boldsymbol{f}}(\bb{x}_i,y_i)\big]\bigg]\\
  =&\mathbb{E}_{\bb{\sigma}} \bigg[\sup _{\bb{f} \in \mathcal{F}} \frac{1}{m}\sum_{i=1}^m \sigma_i \frac{\bb{f}_y(\bb{x}_i)-\underset{y'\neq y_i}{\max}\bb{f}_{y'}(\bb{x}_i)}{2\mu}\bigg]\\
  \leq& \frac{1}{2\mu}\underbrace{\mathbb{E}_{\bb{\sigma}} \bigg[\sup _{\bb{f} \in \mathcal{F}} \frac{1}{m}\sum_{i=1}^m \sigma_i \bb{f}_{y_i}(\bb{x}_i)\bigg]}_{:=\widehat{\mathfrak{R}}_{{\widehat{\mathcal{D}}}}(\Omega_1(\mathcal{F}))}+\\
  &\frac{1}{2\mu}\underbrace{\mathbb{E}_{\bb{\sigma}} \bigg[\sup _{\bb{f} \in \mathcal{F}} \frac{1}{m}\sum_{i=1}^m \sigma_i \underset{y'\neq y_i}{\max}\bb{f}_{y'}(\bb{x}_i)\bigg]}_{:=\Delta}.
\end{aligned}
\end{equation}

Because $\mathbb{E}_{\bb{\sigma}} \big[\sup _{\bb{f} \in \mathcal{F}} \frac{1}{m}\sum_{i=1}^m \sigma_i \mathbb{W}_\mu\big(v)\big]= 0$ holds for $\mu\leq v$ or $v\leq 0$, so here we only consider the case of $0\leq v\leq \mu$. Let $\Omega_1(\mathcal{F}^{\mathcal{C}-1})=\{\max\{\bb{f}_1,...,\bb{f}_{\mathcal{C}-1}\}~|~\bb{f}_j \in \Omega_1(\mathcal{F}), j\in \{1,...,\mathcal{C}-1\}\}$. Then according to Lemma \ref{lem_3}, the term $\Delta$ can be rewritten as:
\begin{equation}
\begin{aligned}
    \label{a22}
    \Delta&=\mathbb{E}_{\bb{\sigma}} \bigg[\sup _{\bb{f} \in \mathcal{F}} \frac{1}{m}\sum_{i=1}^m \sigma_i \underset{y'\neq y_i}{\max}\bb{f}_{y'}(\bb{x}_i)\bigg]\\
    &=\mathbb{E}_{\bb{\sigma}} \bigg[\sup _{\bb{f} \in \mathcal{F}} \frac{1}{m}\sum_{i=1}^m \sigma_i \underset{j\in \{1,...,\mathcal{C}-1\}}{\max}\bb{f}_{j}(\bb{x}_i)\bigg]\\
    &=\widehat{\mathfrak{R}}_{{\widehat{\mathcal{D}}}}(\Omega_1(\mathcal{F}^{\mathcal{C}-1}))\\
    &\leq (\mathcal{C}-1)\widehat{\mathfrak{R}}_{{\widehat{\mathcal{D}}}}(\Omega_1(\mathcal{F})).
\end{aligned}
\end{equation}

Therefore, $ \widehat{\mathfrak{R}}_{{\widehat{\mathcal{D}}}}(\mathcal{G})\leq\frac{\mathcal{C}}{2\mu}\widehat{\mathfrak{R}}_{{\widehat{\mathcal{D}}}}(\Omega_1(\mathcal{F}))$. This completes the proof of Proposition \ref{pro_2}.
\vspace{10pt}

\begin{proposition}
    \label{pro_4}
    Let $\mathcal{G}$ be a family of margin loss functions defined as $\mathbb{E}_{\bb{x}\sim \mathcal{D}_x}\big\{\mathbb{W}_\mu\big[\omega_{\bb{f'}}\big(x,h_{\bb{f}}(x)\big)\big]\big\}$ and $\Omega_2(\mathcal{F})$ be a family of functions defined in Definition \ref{def_2}. Then, \tx{for any sample $\widehat{\mathcal{D}}_x=\{\bb{x}_i\}_{i=1}^{m}$}, the following relation holds between the empirical Rademacher complexities of $\mathcal{G}$ and $\Omega_2(\mathcal{F})$:
    \begin{equation}
    \widehat{\mathfrak{R}}_{{\widehat{\mathcal{D}}}_x}(\mathcal{G})\leq\frac{\mathcal{C}}{2\mu}\widehat{\mathfrak{R}}_{{\widehat{\mathcal{D}}_x}}(\Omega_2(\mathcal{F})).
    \end{equation}
\end{proposition}

\textit{\textbf{Proof.}} The proof is similar to Proposition \ref{pro_2}, the empirical Rademacher complexity of $\mathcal{G}$ can be written as:
\begin{equation}
\begin{aligned}
    \label{a21}
		&\widehat{\Re}_{\widehat{\mathcal{D}}_x}(\mathcal{G})\\
  =& \mathbb{E}_{\bb{\sigma}} \bigg[\sup _{\bb{f},\bb{f'} \in \mathcal{F}} \frac{1}{m}\sum_{i=1}^m \sigma_i \mathbb{W}_\mu\big[\omega_{\boldsymbol{f'}}(\bb{x}_i,\bb{f}(x))\big]\bigg]\\
  =&\mathbb{E}_{\bb{\sigma}} \bigg[\sup _{\bb{f},\bb{f'} \in \mathcal{F}} \frac{1}{m}\sum_{i=1}^m \sigma_i \frac{\bb{f'}_{h_{\bb{f}}(\bb{x})}(\bb{x}_i)-\underset{y'\neq h_{\bb{f}}(\bb{x})}{\max}\bb{f}_{y'}(\bb{x}_i)}{2\mu}\bigg]\\
  \leq& \frac{\mathcal{C}}{2\mu}\widehat{\mathfrak{R}}_{{\widehat{\mathcal{D}}_x}}(\Omega_2(\mathcal{F})).
\end{aligned}
\end{equation}

This completes the proof of Proposition \ref{pro_4}.
\vspace{10pt}

\begin{proposition}\label{pro_3}
    Let $\widehat{P}$ and $\widehat{Q}$ be the corresponding empirical distributions for sample $\mathcal{D}_{s}=(\bb{X},\bb{Y}_j)$ and $\mathcal{D}_t=\bb{X}_t$. For any $\delta>0$, with probability at least $1-2\delta$, the following holds for any scoring function $\bb{f} \in \mathcal{F}$,
    \begin{equation*}
    \begin{aligned}
         \left|d^{\mu}_{\bb{f}}\left(P, Q\right)-d^{\mu}_{\bb{f}}\left(\widehat{P}, \widehat{Q}\right)\right|\leq\frac{\mathcal{C}}{\mu}\widehat{\mathfrak{R}}_{{\widehat{\mathcal{D}}_s}}(\Omega_2(\mathcal{F}))+3\sqrt{\frac{\log (2 / \delta)}{2|\widehat{\mathcal{D}}_s|}}\\
         +\frac{\mathcal{C}}{\mu}\widehat{\mathfrak{R}}_{{\widehat{\mathcal{D}}_t}}(\Omega_2(\mathcal{F}))+3\sqrt{\frac{\log (2 / \delta)}{2|\widehat{\mathcal{D}}_t|}}.
    \end{aligned}
    \end{equation*}
\end{proposition}

\textbf{\textit{Proof.}} Eq. \eqref{5} gives the margin disparity discrepancy $	d^{\mu}_{\bb{f}}\left(P, Q\right)=\sup _{\boldsymbol{f}^{\prime} \in \mathcal{F}}\Big\{\mathbb{E}_{Q}\big[ \mathbb{W}_\mu\left(\omega_{\boldsymbol{f}^{\prime}}\right)\big]-\mathbb{E}_{P} \big[\mathbb{W}_\mu\left(\omega_{\boldsymbol{f}^{\prime}}\right)\big]\Big\}$, then we have,
\begin{equation}
\label{1112}
    \begin{aligned}
        &\left|d^{\mu}_{\bb{f}}\left(P, Q\right)-d^{\mu}_{\bb{f}}\left(\widehat{P}, \widehat{Q}\right)\right|\\
        \leq& \sup _{\boldsymbol{f}^{\prime} \in \mathcal{F}}\Big|\mathbb{E}_{Q}\big[ \mathbb{W}_\mu\left(\omega_{\boldsymbol{f}^{\prime}}\right)\big]-\mathbb{E}_{\widehat{Q}} \big[\mathbb{W}_\mu\left(\omega_{\boldsymbol{f}^{\prime}}\right)\big]\Big|\\
        +& \sup _{\boldsymbol{f}^{\prime} \in \mathcal{F}}\Big|\mathbb{E}_{P}\big[ \mathbb{W}_\mu\left(\omega_{\boldsymbol{f}^{\prime}}\right)\big]-\mathbb{E}_{\widehat{P}} \big[\mathbb{W}_\mu\left(\omega_{\boldsymbol{f}^{\prime}}\right)\big]\Big|.
    \end{aligned}
\end{equation}

By applying Lemma \ref{lem_2} and Proposition \ref{pro_4}, then
\begin{equation}
\label{1111}
    \begin{aligned}
        &\Big|\mathbb{E}_{Q}\big[ \mathbb{W}_\mu\left(\omega_{\boldsymbol{f}^{\prime}}\right)\big]-\mathbb{E}_{\widehat{Q}} \big[\mathbb{W}_\mu\left(\omega_{\boldsymbol{f}^{\prime}}\right)\big]\Big|\\
        \leq&\frac{\mathcal{C}}{\mu}\widehat{\mathfrak{R}}_{{\widehat{\mathcal{D}}}_t}(\Omega_2(\mathcal{F}))+3\sqrt{\frac{\log (2 / \delta)}{2|\widehat{\mathcal{D}_t}|}},\\
        &\Big|\mathbb{E}_{P}\big[ \mathbb{W}_\mu\left(\omega_{\boldsymbol{f}^{\prime}}\right)\big]-\mathbb{E}_{\widehat{P}} \big[\mathbb{W}_\mu\left(\omega_{\boldsymbol{f}^{\prime}}\right)\big]\Big|\\
        \leq&\frac{\mathcal{C}}{\mu}\widehat{\mathfrak{R}}_{{\widehat{\mathcal{D}}}_s}(\Omega_2(\mathcal{F}))+3\sqrt{\frac{\log (2 / \delta)}{2|\widehat{\mathcal{D}_s}|}}.
    \end{aligned}
\end{equation}

Combing Eqs. \eqref{1112} and \eqref{1111}, we can get the final result, which completes the proof of Proposition \ref{pro_3}. Next, we give the proof for Theorem \ref{the_2}.
\vspace{10pt}

{\textbf{The Proof for Theorem \ref{the_2}}}: First, according to Lemma \ref{lem_2} and Proposition \ref{pro_2}, for $\mathcal{L}_{P}^{\mu}(\bb{f})=\mathbb{E}_{(\bb{x},y)\sim P}\big\{ \mathbb{W}_\mu\big[\omega_{\boldsymbol{f}}(\bb{x},y)\big]\big\}$, we have
\begin{equation}
    \left|\mathcal{L}_{{P}}^{\mu}\left(\bb{f}\right)-\mathcal{L}_{\widehat{P}}^{\mu}\left(\bb{f}\right)\right|\leq \frac{\mathcal{C}}{\mu}\widehat{\mathfrak{R}}_{{\widehat{\mathcal{D}}}}(\Omega_1(\mathcal{F}))+3\sqrt{\frac{\log (2 / \delta)}{2|\widehat{\mathcal{D}}|}}.
\end{equation}

Then, combining Theorem \ref{the_1} and Proposition \ref{pro_3}, we have 
	\begin{equation}
		\begin{aligned}
			&\mathcal{L}_Q\left(h_{\bb{f}}\right) \leq \sum_{j=1}^N \alpha_j\bigg\{\mathcal{L}_{{\widehat{P}}_j}^{\mu}\left(\bb{f}\right)+d^\mu_{\bb{f}}\left(\widehat{P}_j, \widehat{Q}\right)\\
			+&\frac{\mathcal{C}}{\mu}{\widehat{\mathfrak{R}}}_{\widehat{\mathcal{D}}_{sj}}(\Omega_2(\mathcal{F}))+\frac{\mathcal{C}}{\mu}{\widehat{\mathfrak{R}}}_{\widehat{\mathcal{D}}_{sj}}(\Omega_1(\mathcal{F}))+6\sqrt{\frac{\log (2 / \delta)}{2 |\widehat{\mathcal{D}}_{sj}|}}\bigg\}\\
			+&\frac{\mathcal{C}}{\mu}{\widehat{\mathfrak{R}}}_{\widehat{\mathcal{D}}_{t}}(\Omega_2(\mathcal{F}))+3\sqrt{\frac{\log (2 / \delta)}{2 |\widehat{\mathcal{D}}_{t}|}}+\beta.
		\end{aligned}
	\end{equation}

Since $\widehat{\mathcal{D}}_{sj}$ and $\widehat{\mathcal{D}}_t$ are empirical datasets sampled from the i.i.d. sample $\mathcal{D}_{sj}=(\bb{X}_j,\bb{Y}_j)$ and $\mathcal{D}_t=\bb{X}_t$ of size $m$, the above equation can be written as
	\begin{equation*}
		\begin{aligned}
			\mathcal{L}_Q\left(h_{\bb{f}}\right) \leq &\sum_{j=1}^N \alpha_j\Big\{\mathcal{L}_{{\widehat{P}}_j}^{\mu}\left(\bb{f}\right)+d^\mu_{\bb{f}}\left(\widehat{P}_j, \widehat{Q}\right)\Big\}+9\sqrt{\frac{\log (2 / \delta)}{2m}}\\
			&+\frac{\mathcal{C}}{\mu}\Big\{{\widehat{\mathfrak{R}}}_{\widehat{\mathcal{D}}_s}(\Omega_1(\mathcal{F}))+{\widehat{\mathfrak{R}}}_{\widehat{\mathcal{D}}}(\Omega_2(\mathcal{F}))\Big\}+\beta,
		\end{aligned}
	\end{equation*}
where $\widehat{\mathcal{D}}_s$ and $\widehat{\mathcal{D}}$ are datasets sampled from the mixture distribution $\sum_{j=1}^{N}\alpha_jP_j$ and $Q+\sum_{j=1}^{N}\alpha_jP_j$, respectively. \tc{This completes the proof of Theorem \ref{the_2}.}

\subsection{Proof of Theorem \ref{the_4}}\label{proof_the_4}
\textbf{{The Proof for Theorem \ref{the_4}}}: This proof follows the previous works \cite{goodfellow2014generative,zhang2019bridging}. Since the function $f(y)=a\log(y)+b\log(1-y)$ achieves its maximum in $[0,1]$ at $a/(a+b)$, for the below optimization problem: $\max \gamma\mathbb{E}_{P}[\log D(\bb{x})]+\mathbb{E}_{Q}[\log (1- D(\bb{x}))]$, the optimal $D(\bb{x})$ is
\begin{equation}
    D^*(\bb{x})=\frac{\gamma P(\bb{x})}{\gamma P(\bb{x})+Q(\bb{x})}.
\end{equation}

Then, under the optimal $D^*(\bb{x})$, the maximization of $\gamma\mathbb{E}_{P}[\log D(\bb{x})]+\mathbb{E}_{Q}[\log (1- D(\bb{x}))]$ can be denoted as
\begin{equation}
\begin{aligned}
    &\gamma\mathbb{E}_{P}\big[\log D(\bb{x})\big]+\mathbb{E}_{Q}\big[\log (1- D(\bb{x}))\big]\\
    =&\int_{\bb{x}\in\mathcal{X}} \gamma P(\bb{x})\log\Big[\frac{\gamma P(\bb{x})}{\gamma P(\bb{x})+Q(\bb{x})}\Big] \mathrm{d}x\\
    +&\int_{\bb{x}\in\mathcal{X}}Q(\bb{x})\log\Big[\frac{Q(\bb{x})}{\gamma P(\bb{x})+Q(\bb{x})}\Big] \mathrm{d}x\\
    =&\int_{\bb{x}\in\mathcal{X}} \gamma P(\bb{x})\left\{\log\Big[\frac{\gamma P(\bb{x})}{\frac{\gamma P(\bb{x})+Q(\bb{x})}{\gamma+1}}\Big]-\log\left[\frac{\gamma+1}{\gamma}\right]\right\}\mathrm{d}x \\
    +& \int_{\bb{x}\in\mathcal{X}}Q(\bb{x})\left\{\log\Big[\frac{Q(\bb{x})}{\frac{ P(\bb{x})+Q(\bb{x})}{\gamma+1}}\Big]-\log\left[\gamma+1\right]\right\}\mathrm{d}x\\
    =&\int_{\bb{x}\in\mathcal{X}} \gamma P(\bb{x})\log\Big[\frac{P(\bb{x})}{\frac{\gamma P(\bb{x})+Q(\bb{x})}{\gamma+1}}\Big]\mathrm{d}x \\
    +& \int_{\bb{x}\in\mathcal{X}}Q(\bb{x})\log\Big[\frac{Q(\bb{x})}{\frac{\gamma P(\bb{x})+Q(\bb{x})}{\gamma+1}}\Big]\mathrm{d}x\\
    +&\int_{\bb{x}\in\mathcal{X}} \gamma P(\bb{x}) \log\gamma -\big[\gamma P(\bb{x})+Q(\bb{x}) \big] \log (\gamma+1)\mathrm{d}x\\
    =&\gamma KL\Big(P ~\Big|\Big|~\frac{\gamma P+Q}{\gamma+1}\Big)+KL\Big(Q ~\Big|\Big|~\frac{\gamma P+Q}{\gamma+1}\Big)\\
    &+\gamma \log\gamma-(\gamma+1)\log(\gamma+1).
\end{aligned}
\end{equation}

Therefor, when the above $D(\bb{x})$ represents the function $\Theta_{h_{\bb{f'}_j}}\big(\bb{f}(\phi(\bb{x}))\big)$, fixing the classifier $\bb{f}$, the theorem \ref{the_4} can be proven.

\end{document}